\documentclass[11pt,a4paper]{article}
\usepackage[utf8]{inputenc}
\usepackage{textcomp}
\usepackage{graphicx,verbatim,array,multicol,courier}
\usepackage{amsmath,amssymb,amsthm,mathrsfs, mathtools}
\usepackage{color, graphics, graphicx, siunitx}
\usepackage{amsfonts, dsfont, bm}
\usepackage{natbib} 
\usepackage{a rydshln}
\usepackage{geometry}
\usepackage{booktabs}
\usepackage[normalem]{ulem}
\usepackage{enumitem,float}
\usepackage{lscape}
\usepackage{dsfont}
\usepackage{textcomp}
\usepackage[pdftex,bookmarks,colorlinks]{hyperref}
\usepackage[dvipsnames]{xcolor}
\usepackage{tabularx}
\usepackage{siunitx}
\usepackage{eurosym}
\usepackage{siunitx}
\usepackage[labelformat=simple]{subfig}
\usepackage[linesnumbered,ruled,vlined]{algorithm2e}
\usepackage[toc,page]{appendix}
\usepackage{booktabs} 

\usepackage{arydshln}
\usepackage{makecell}
\usepackage{booktabs,arydshln}
 \usepackage{multirow}
 \usepackage{textcomp} 
\usepackage{siunitx}  
\DeclareSIUnit\permille{\text{\textperthousand}}

\newlist{inparaenum}{enumerate}{2}
\setlist[inparaenum,1]{label=(\alph*)}
\setlist[inparaenum,2]{label=(\roman{inparaenumi}\emph{\alph*})}
\makeatletter
\def\adl@drawiv#1#2#3{%
        \hskip.5\tabcolsep
        \xleaders#3{#2.5\@tempdimb #1{1}#2.5\@tempdimb}%
                #2\z@ plus1fil minus1fil\relax
        \hskip.5\tabcolsep}
\newcommand{\cdashlinelr}[1]{%
  \noalign{\vskip\aboverulesep
           \global\let\@dashdrawstore\adl@draw
           \global\let\adl@draw\adl@drawiv}
  \cdashline{#1}
  \noalign{\global\let\adl@draw\@dashdrawstore
           \vskip\belowrulesep}}
\makeatother

\numberwithin{equation}{section}
\theoremstyle{definition}

\theoremstyle{plain}

\theoremstyle{remark}

\theoremstyle{example}

\newcommand{\diff}{\mathrm{d}}

\definecolor{navy}{rgb}{0,0,0.502}
\definecolor{brown}{rgb}{0.59, 0.29, 0.0}
\def\indic{\mathds{1}}

\hypersetup{colorlinks,%
	citecolor=blue,%
	filecolor=green,%
	linkcolor=red,%
	urlcolor=violet,%
}

\newcommand{\Real}{\mathbb{R}}

\newcommand{\Prob}{\mathbb{P}}

\newcommand{\bftheta}{{\boldsymbol{{\theta}}}}

\newcommand{\bfX}{{\boldsymbol{X}}}
\newcommand{\bfY}{{\boldsymbol{Y}}}
\newcommand{\bfZ}{{\boldsymbol{Z}}}

\newcommand{\bfy}{{\boldsymbol{y}}}

\newcommand{\bfx}{{\boldsymbol{x}}}

\usepackage[font=small, textfont=it]{caption}

\title{Predicting hazards of climate extremes: a statistical perspective}
\author{Carlotta Pacifici\footnote{BAFFI - Centre on Economics, Finance and Regulation, Bocconi University, Italy. E-mail: carlotta.pacifici@unibocconi.it. Orcid: https://orcid.org/0009-0005-6604-2427.}, Simone A. Padoan\footnote{Department of Decision Sciences, Bocconi University, Italy. E-mail: simone.padoan@unibocconi.it. Orcid: https://orcid.org/0000-0002-0417-7570} and Jaroslav Mysiak\footnote{CMCC Foundation - Euro-Mediterranean Center on Climate Change, Italy, and Venice Ca' Foscari University, Italy. E-mail: jaroslav.mysiak@cmcc.it. Orcid: https://orcid.org/0000-0001-9341-7048}}

\begin{document}
\maketitle
\begin{abstract}
Climate extremes 
such as floods, storms, and heatwaves have caused severe economic and human losses across Europe in recent decades. To support the European Union’s climate resilience efforts, we propose a statistical framework for short-to-medium-term prediction of tail risks related to extreme economic losses and fatalities. Our approach builds on Extreme Value Theory and employs the predictive distribution of future tail events to quantify both estimation and aleatoric uncertainty. Using data on EU-wide losses and fatalities from 1980 to 2023, we model extreme events through Peaks Over Threshold methodology and fit Generalised Pareto (GP) and discrete-GP models using an empirical Bayes procedure. Our predictive approach enables a “What-if” analysis to evaluate hypothetical scenarios beyond observed levels, including potential worst-case outcomes for a precautionary risk assessment of future extreme episodes. To account for a time-varying behavior of extreme losses and fatalities we extend our predictive method using a proportional tail model that allows to handle heteroscedastic extremes over time. 
Results of our analysis under stationarity and non-stationary settings raise concerns, 
reinforcing the urgency of integrating predictive tail risk assessment into EU adaptation strategies.
\end{abstract}
\section{Introduction}\label{sec:intro}
%
%
\subsection{European climate risk management objectives}\label{sec:motivation}
%

Climate-related hazards such as floods, storms, and heatwaves pose significant threats to ecosystems, infrastructure, and public health, often leading to severe economic losses and fatalities. In recent years, Europe has faced numerous extreme events with catastrophic consequences: the 2021 floods in Belgium and Germany caused together a total amount of \euro47 billion in damage and over 200 deaths \citep{newsfloodDE2021}, while the 2003 heatwave resulted in \euro20 billion in losses and nearly 75,000 fatalities—making the summer of 2003 the hottest in Europe at that time since 1500 \citep{luterbacher2004european}.

To address these risks, the European Union (EU) revised the EU Adaptation Strategy in 2021, 
%
aiming for a climate-resilient society through smart, fast, and systematic adaptation. 
The European Climate Risk Assessment (EUCRA) identifies five key climate risk clusters, highlighting human health and economic stability as particularly critical. It warns of severe risks, such as heat stress and financial market instability, and calls for urgent action. 
EU policies face two major challenges. 
First, while most adaptation strategies are designed for medium-to long term, climate risks are evolving at a pace that increasingly challenges the speed of policy implementation 
 \citep[pages 12 and 41]{newsEEA}. 
%
Second, current policies are ill-suited to address the full spectrum of climate risks, as they tend to focus on average scenarios while overlooking low-probability, high-impact “tail risks” \citep[page 41]{newsEEA}.
The EEA stresses the urgent need for precautionary risk management to address extreme scenarios with potentially catastrophic consequences. Effective adaptation and mitigation policies must account for tail risks to enhance Europe’s resilience against future climate-related disasters \citep{newsEEAindicator,newsEEA}.
To support EU resilience efforts,  {we propose a statistical analysis of historical data on climate-related losses and fatalities} for short-to medium term prediction of “tail risks”, whose outcome can be used for the implementation of adaptation strategies.
%
\subsection{Europe’s climate losses and fatalities data}\label{sec:data}
%
%
\begin{figure}[t!]
\begin{center}
\includegraphics[width=0.25\textwidth, page=6]{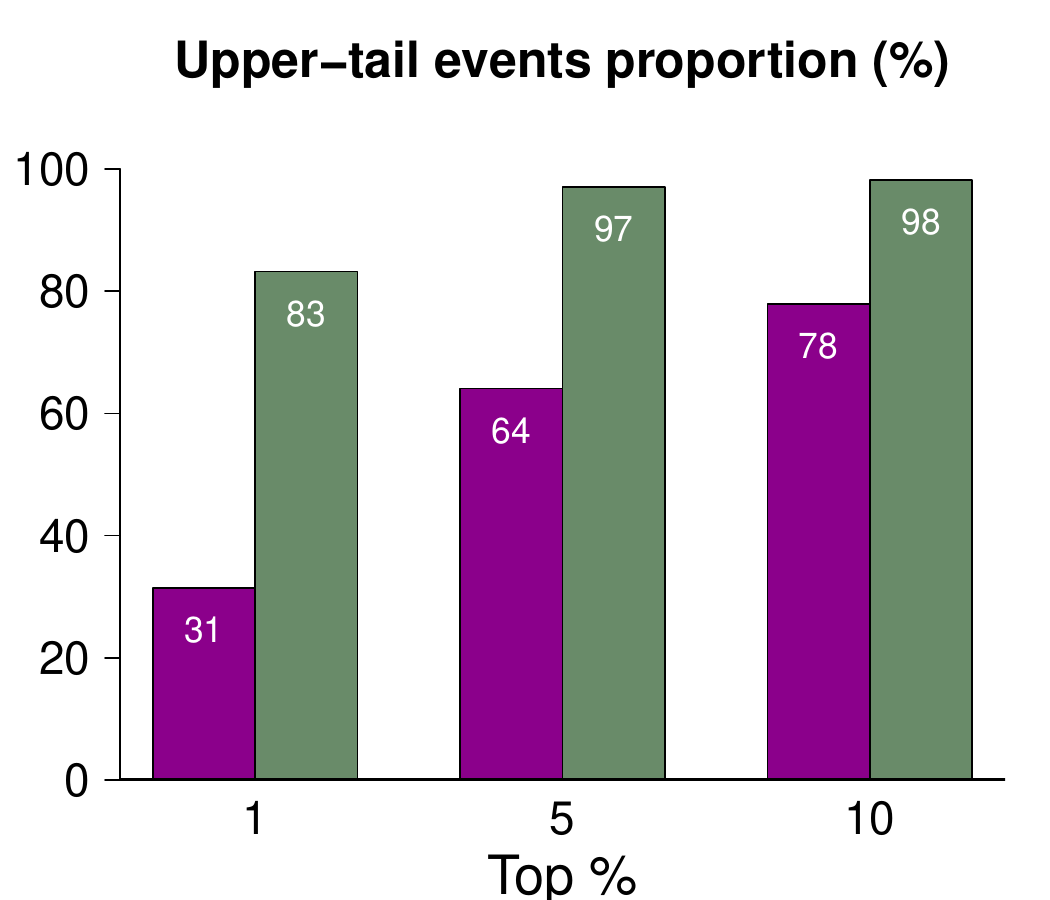}
\includegraphics[width=0.25\textwidth, page=7]{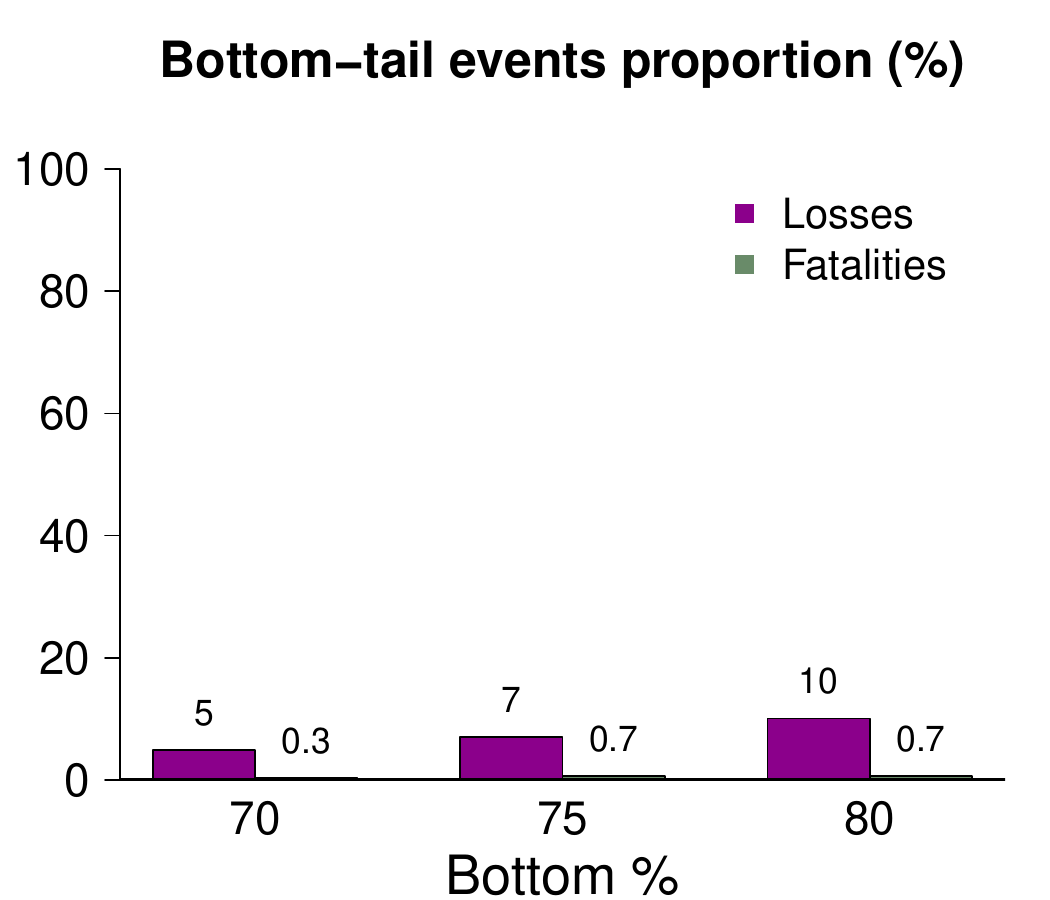}\\
\includegraphics[width=0.25\textwidth, page=8]{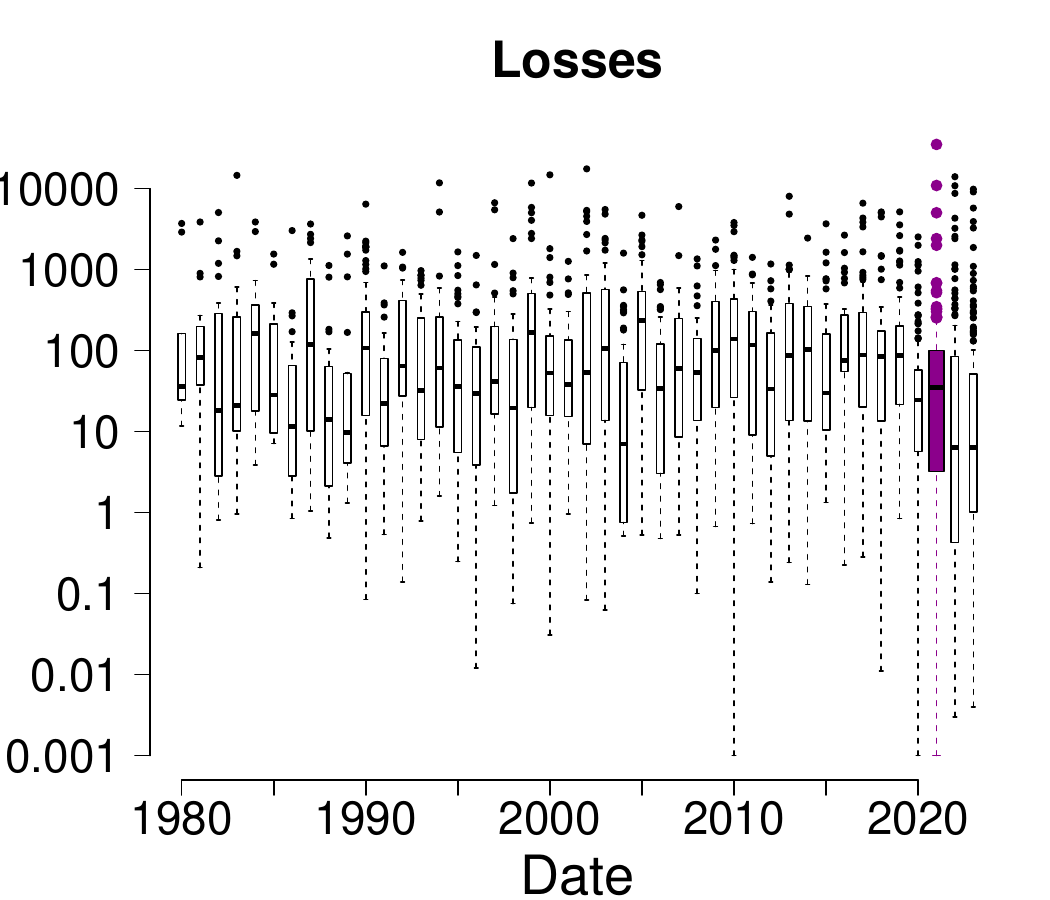}
\includegraphics[width=0.25\textwidth, page=9]{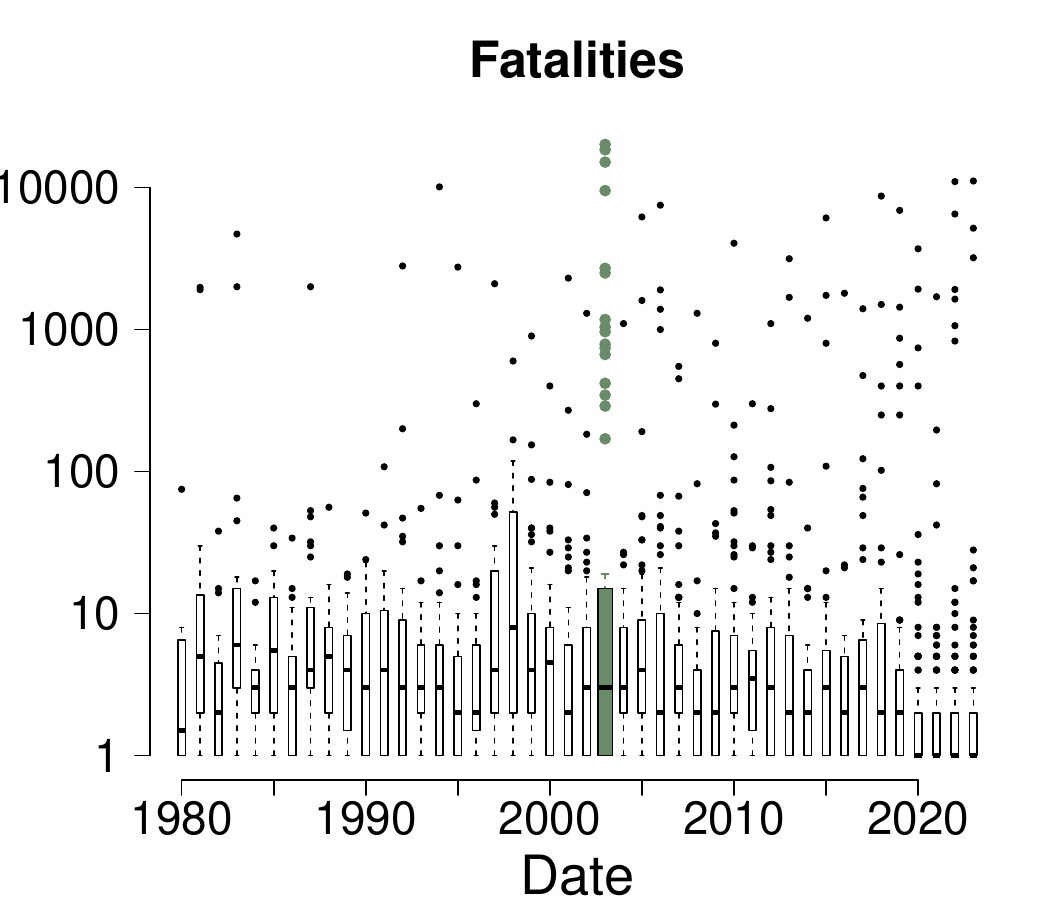}\\
\includegraphics[width=0.25\textwidth, page=2]{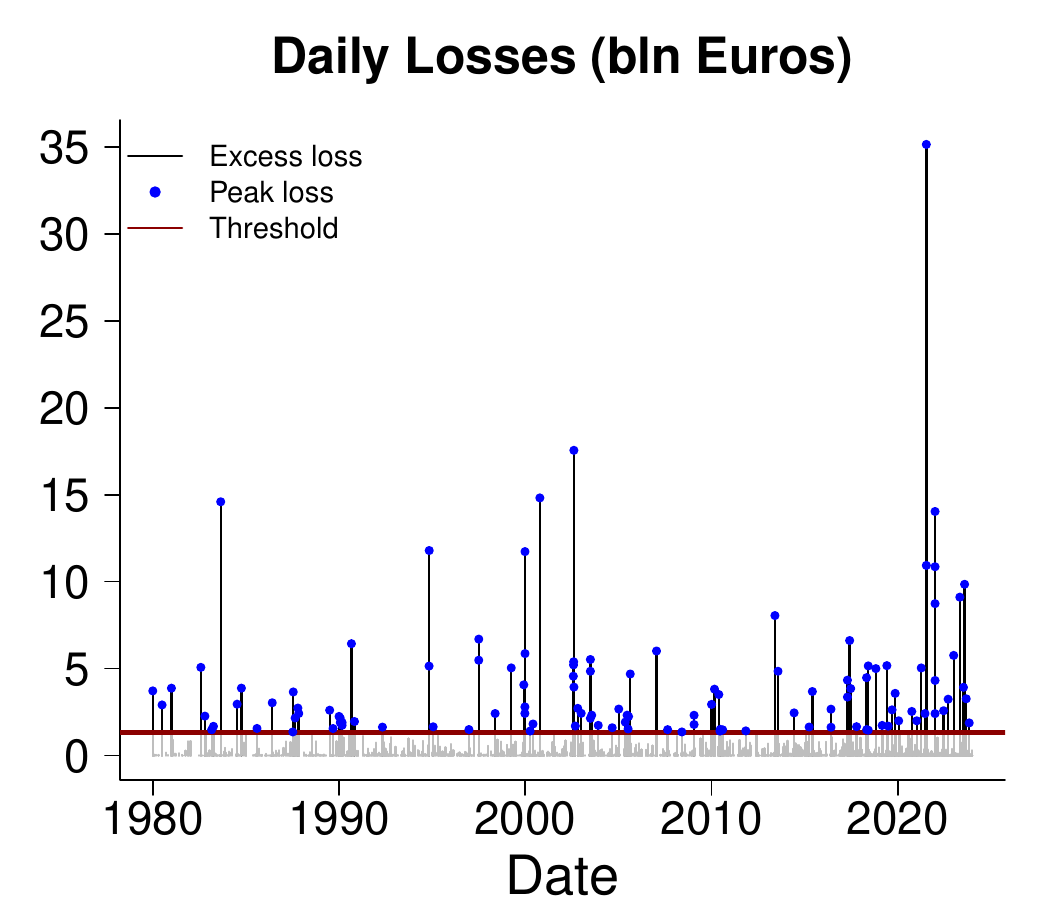}
\includegraphics[width=0.25\textwidth, page=3]{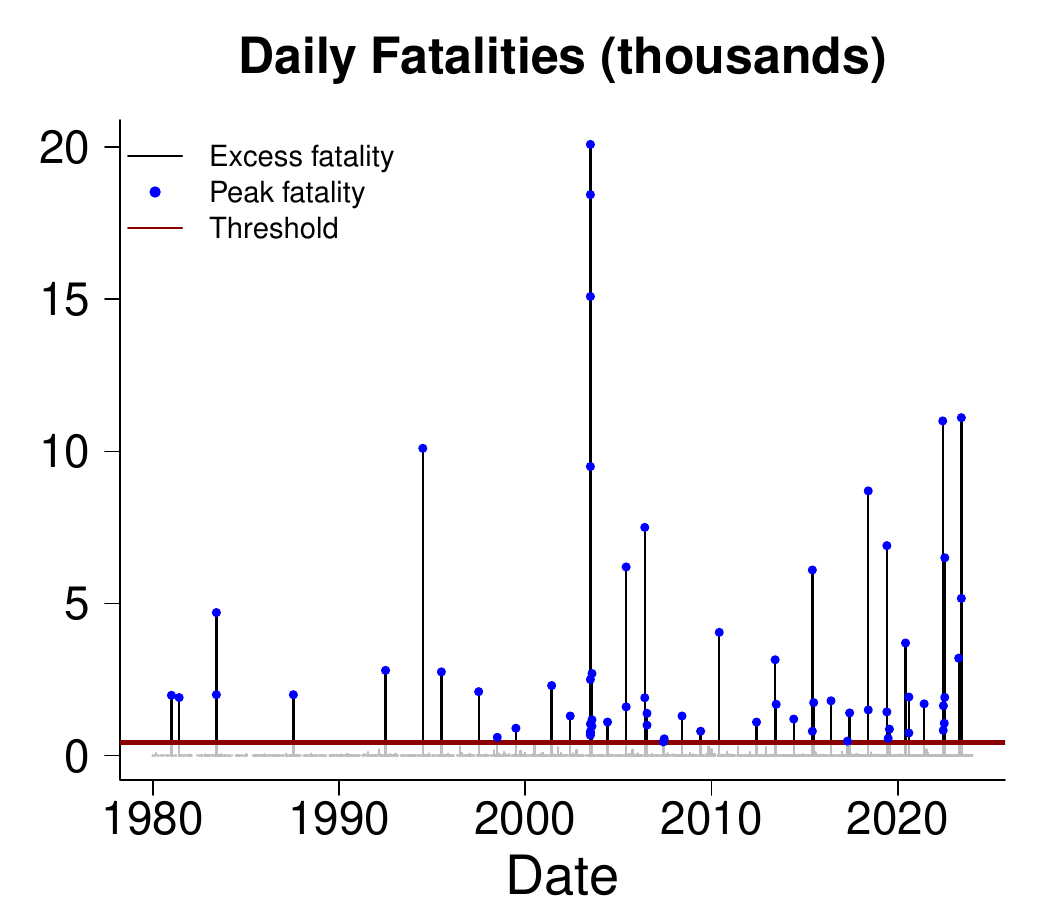}\\
\includegraphics[width=0.25\textwidth, page=4]{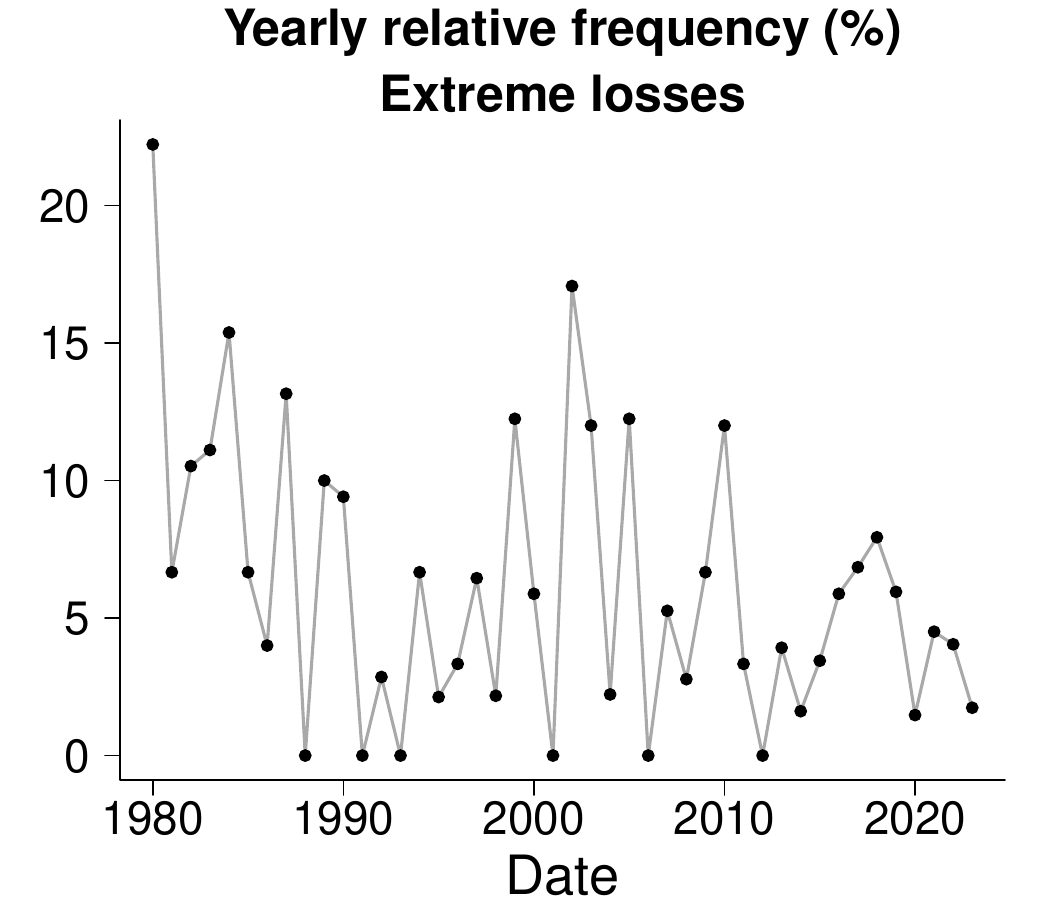}
\includegraphics[width=0.25\textwidth, page=5]{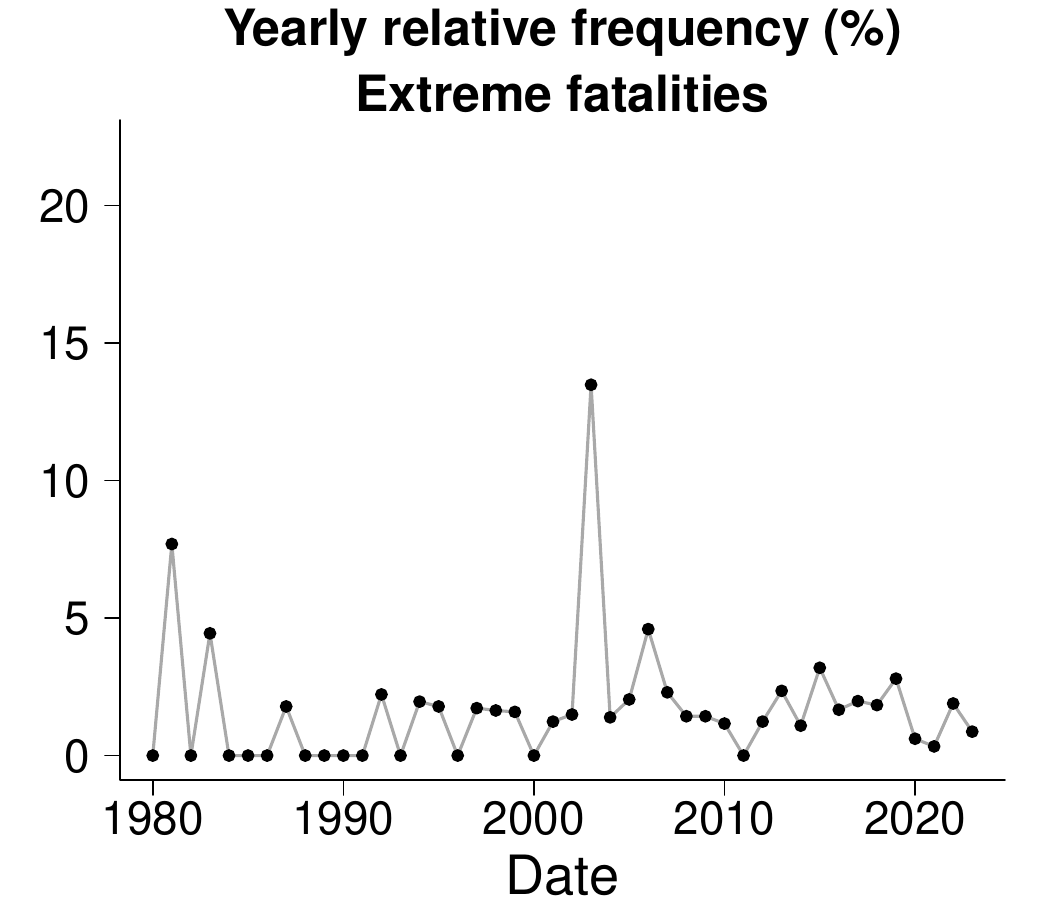}
\caption{ 
Top panels show the share of total losses (dark magenta) and fatalities (light green) contributed by the top 1\%, 5\%, 10\% and bottom 70\%, 75\%, 80\% of events. Those in the second row display the yearly distribution of  {positive} daily losses and fatalities by event type (log scale), with colored boxplots marking the years of the highest recorded values. Panels in the third row show losses (top-left) and fatalities (top-right) in gray dots with peaks in blue above a high threshold (red line). Panels in fourth row display the proportion of loss and fatality events, respectively, that exceed a high threshold relative to the total number of corresponding events.}
\label{fig:bubbles}
\end{center}
\end{figure}
%
We analyse economic losses (losses) and fatalities from floods, storms, and extreme temperatures across  {the} EU-27 countries from 1980 to 2023, 
for a total of 4,055 observations. 
The data were obtained from the Catastrophe Database (CATDAT) by RiskLayer (\url{https://www.risklayer.com}). 

The top panels of Figure \ref{fig:bubbles} examine loss and fatality distributions, with the left one showing that tail events dominate total losses—for instance, the top 5\% of losses account for 64\% of the total, while the bottom 75\% represent only 7\%.  A similar pattern is observed for fatalities,  {suggesting} the likelihood of heavy-tailed distributions. The panels in the second row of Figure \ref{fig:bubbles} confirm the presence of  heavy-tailed feature across most years, with colored boxplots highlighting the years with the highest daily loss and fatality records. The largest loss recorded in a particular day is approximately \euro35 billion, occurred in Germany in July 2021, while the deadliest day, with 20,089 fatalities, was recorded in Italy in July, 2003.
The panels in the third row of Figure \ref{fig:bubbles} present the time series of losses (left) and fatalities (right), with peaks exceeding the 95th and 98th quantiles highlighted in blue. The panels in the last row  {show} a decreasing share over time of extreme losses over the yearly total number of losses, whereas extreme fatalities exhibit strong fluctuations, with a decreasing trend from approximately 2003 to 2018 and after an increasing trend until 2023.  The observed trend in losses may be explained as follows: although the absolute frequency of annual peaks remains relatively stable, the total number of annual loss events is increasing. This rise is likely due to growing public awareness and improved reporting of climate-related damages, as well as a significant uptick in the number of medium-to-low severity loss events over time. We consider these features in Section \ref{sec:res:nonstat}.

Given these empirical findings, we adopt the following modeling approach. First, due to the disproportionate impact of extreme events, we focus on peak losses and fatalities at the daily level, rather than their yearly total amount. Second, we do not aggregate the data by country and type of event (heatwaves, floods, etc.) but analyse all the distinct events together. This allows us to see them as almost independent events. Third, we opt for independent marginal analysis for simplicity. While this is a simplification, it allows for more accurate statistical predictions of extreme events. Indeed, a joint analysis would introduce two major challenges: (i) handling complex extremal dependence structures (\citealp[][Ch. 8]{coles2001}; \citealp[][Ch. 6]{haan2006extreme}; \citealp{beranger2015extreme}) which make inference complicated \citep[][]{huser2016likelihood, marcon2017, hanson2017bernstein, vettori2018comparison} and make prediction an open problem, and (ii) the methodological gap in modeling joint extremes of a continuous (losses) and a discrete (fatalities) variable. There is a reason for analysing mortality and economic damage separately—early warning systems have improved over the past decades, which helps to protect lives but cannot prevent damages beyond a certain threshold.

%
\subsection{Main contributions and paper outline}\label{sec:contribution}
%

We propose a statistical analysis of data on climate-related losses and fatalities with the aim of predicting the related tail risks, which can help EU in the development of adaptation strategies.
Our approach leverages Extreme Value Theory (EVT, \citealp[e.g.,][]{coles2001, haan2006extreme, de2016statistics}) and in particular the Peaks Over Threshold (POT, \citealp{davisonsmith1990}) method, widely used for analysing peaks using the Generalized Pareto (GP) distribution \citep{balkema1974residual} as an approximate model. While traditionally applied to continuous variables, we also discuss the case of discrete outcomes, such as fatalities, using the discrete-Generalized Pareto (d-GP) distribution \citep[e.g.,][]{shimura2012discretization,chavez2022dGP, koh2023gradient, hitz_davis_samorodnitsky_2024}. 

In the first step, we fit the peak losses (or fatalities) to the GP (d-GP) model leveraging the empirical Bayes procedure of \citet{dombry2023asymptotic}—a method based on a class of data-dependent prior distributions and similarly applied to block maxima by \citet{padoan2024empirical}. This approach fulfills certain theoretical properties that guarantee highly accurate inference of tail heaviness, offering a robust tool that we employ here for monitoring the severity and potential impact of extreme losses and fatalities (see \citealp{vettori2019bayesian}, \citealp{beranger2021estimation}, \citealp{lee2024bayesian}, \citealp{de2025semiparametric}, for related works on Bayesian methods for extremes).

In classical literature, the most widely used approach for tail risk assessment involves extrapolating risk measures (e.g., extreme quantile, expected shortfall, or expectile; see, e.g., \citealp[][Ch. 4]{haan2006extreme}, \citealp{mcneil2015quantitative}, \citealp{davison2023tail}),  that estimate the severity of an event with a very low probability of exceedance. While this method offers clear advantages, it however only captures estimation (epistemic) uncertainty and overlooks aleatoric uncertainty—stemming from the intrinsic randomness and unpredictability of future events within the phenomenon being analysed. Accurate assessment of uncertainty in future tail events is crucial for designing effective adaptation strategies. To this end, we adopt a predictive approach that allows  {to} appropriately incorporate aleatoric uncertainty in unprecedented episode evaluations. In statistics, predictions are achieved by the specification of suitable predictive distributions \citep[e.g.][]{geisser1993predictive}. In a nutshell, given a sample $\bfY_n=(Y_1,\ldots,Y_n)$ of independent and identically distributed (iid) random variables, representing past observations, the statistical target is to infer the distribution $F^\star$ of an out-of-sample variable $Y_{n+1}$, independent of $\bfY_n$, which represents future unobserved events. The inferred $F^\star$ is then used to predict the magnitude of future events, while accounting for their aleatoric uncertainty. Because our interest is predicting future extreme events, we then focus on the distribution of future peaks exceeding a sufficiently large threshold $t$, i.e., $\Prob( Y_{n+1}\leq y \mid  Y_{n+1}>t, \bfY_n)$, and we call it the predictive distribution of tail events. By employing the threshold stability property from EVT \citep[e.g.,][Ch. 1, 5]{falk2010}, we allow the conditioning threshold $t$ to extend deep into the far tail of the data distribution, enabling a comprehensive {\it What-if}  analysis for risk assessment. This approach facilitates the evaluation of hypothetical scenarios beyond observed levels, including potential worst-case outcomes. It  addresses a key question: if an extreme event were to occur tomorrow, how can we estimate today the  possible range of its impact with a certain probability?
The resulting toolbox offers short- to medium-term insights into tail risks, supporting the EEA’s goals of managing uncertainty across key risks in economy and health sectors.

In the second step, we adopt the empirical Bayes framework of \citet{padoan2023statistical} to derive the posterior predictive distribution—obtained by integrating out the GP (d-GP) model parameters with respect to their posterior— that provides a natural method for the estimation of the true predictive distribution of future peaks and a practical tool for their forecasting. \citet{padoan2023statistical} formally show that the posterior predictive density derived from the empirical Bayesian procedure is a highly accurate estimator of the true predictive density.  Its predictive inference satisfies key theoretical properties, ensuring that the resulting forecasts are both reliable and practically useful. Earlier works that have used classical posterior predictive distributions for extreme values prediction are e.g., \cite{smith1999, coles1996bayesian, coles2003, friederichs2012forecast}. Assuming stationarity of extremes, our preliminary analysis shows that if a loss comparable to the worst 1\% of historical events—or even more severe—were to occur, we should expect a loss between approximately \euro5 billion and \euro37 billion with 95\% probability (see Table \ref{tab:pfdist}). This range includes the worst recorded loss of \euro35.15 billion, and, such a loss could plausibly reoccur as early as 2025 (see Figure \ref{fig:uncPoTest}).

Since the frequency of peak losses and fatalities varies over time—likely influenced by climate change, as illustrated in the bottom panels of Figure \ref{fig:bubbles}—in the third step we adopt the proportional tail model proposed by \citet{dombry2023asymptotic}, which is specifically designed to capture time-varying extreme behavior. To this end, we extend the empirical Bayes framework of \citet{padoan2023statistical}, originally developed for stationary extremes, to allow for: (i) inference under the proportional tail model, (ii) modeling the tail distribution of discrete variables such as fatalities, and (iii) forecasting non-stationary extremes.
Even under this more flexible, non-stationary model, the predictions remain concerning. For instance, the average expected loss in 2024 and 2025 is predicted at approximately \euro{ {27} billion and \euro  {30} billion}, respectively. Moreover, with 95\% probability, losses are expected to lie between approximately \euro {3.7}–33 billion in 2024 and \euro4–3{4} billion in 2025.

The remainder of the paper is organized as follows. Our methods form part of the {\bf R} package {\bf ExtremeRisks} available on CRAN. The online supplementary materials contains
technical results. Section \ref{sec:stat_model} presents the stationary and nonstationary statistical models for extreme losses and fatalities. Section \ref{sec:mod:stat:inf} outlines the inferential and predictive procedures, based on an empirical Bayes framework, used in the analysis. Section \ref{sec:results} reports the results of the extreme loss and fatality analysis. Section \ref{sec:conclusion} draws conclusions.

\section{Statistical models}\label{sec:stat_model}
%
%
\subsection{Stationary case}\label{sec:homos}
%

Let $Y$ be a continuous random variable with an unknown distribution function $F$, whose upper-end point is $y_E=\sup( y: F(y)<1)$. We assume $F$ satisfies the domain of attraction condition, in symbols $F \in \mathcal{D}(H_{\gamma})$, with $\gamma\in\Real$. Then, for any $t <y_E$ there exists a scaling function $s(t)>0$ such that the conditional distribution of the rescaled \textit{excess variable} $(Y-t)$, given $Y>t$ satisfies
\begin{equation}\label{eq:EVTtheorem}
\underset{t \to y_E}{\lim} \mathbb{P}\left( \frac{Y-t}{s(t)} \leq z \mid Y > t \right)=H_{\gamma}(z):=  1-\left(1+\gamma z \right)_+^{-\frac{1}{\gamma}},
\end{equation}
for all $z>0$, where $(z)_+=\max(0,z)$ and $H_\gamma$ is the unit-scale Generalised Pareto GP distribution \citep[][Ch. 1]{haan2006extreme}. 
The support of $H_{\gamma}$ is $\mathbb{R}_+$ if $\gamma \geq 0$ and $(-\infty,-1/\gamma)$ if $\gamma <0$. The parameter $\gamma \in \mathbb{R}$ is the \textit{extreme value index}, that describes the heaviness of the tail of $H_{\gamma}$. In the popular book \cite{coles2001} such a parameter is denoted by $\xi$.
$H_{\gamma}$ is heavy-, light- or short-tailed, depending whether $\gamma > 0$, $\gamma=0$ or $\gamma<0$, respectively. Its probability density is $h_{\gamma}(z)= \left( 1+\gamma z \right)_+^{-{1}/{\gamma}-1}$ for $\gamma \neq 0$.
The case $\gamma=0$ leads to $H_{\gamma}(z)=1-\exp(-z)$, for $z>0$,  and $h_{\gamma}(z)=\exp(-z)$, obtained as limit with $\gamma \rightarrow 0$.

The utility of result \eqref{eq:EVTtheorem} is that if we consider the conditional distribution of the \textit{peak variable} $Y$, given that $Y>t$, then  for a high threshold $t$, $\mathbb{P}(Y\leq y\mid Y> t)$ can be approximated by $H_{\bftheta}(y-t) := H_{\gamma}((y-t)/\sigma)$, where $\bftheta=(\gamma,\sigma)^\top$ and the scale parameter $\sigma>0$ is representative of $s(t)$. Its density is $h_{\bftheta}(y-t) =  h_{\gamma}((y-t)/\sigma)/\sigma$.
The strength of EVT lies in its ability to estimate the $\tau$-quantiles of an unknown distribution $F$ for high quantile levesl $\tau$ close to one. These extreme quantiles represent values of the underlying random variable that are expected to be exceeded with only a small probability, $1-\tau<1- F(t)$, making them critical for risk assessment and tail probability estimation. Recall that a $\tau$-quantile is defined as $Q(\tau):=\inf(y:F(y) \geq \tau )$.
Indeed, for large $y>t$, with $t$ close to $y_E$,  \eqref{eq:EVTtheorem} implies 
\begin{align}\label{eq:quant}
F(y)\approx F(t) + (1-F(t))H_{\bftheta}(y-t).
\end{align}
Now, inverting \eqref{eq:quant} we obtain for $\tau\to 1$ the approximation
\begin{equation}\label{eq:quantext}
Q(\tau) \approx t + \sigma\frac{ \left( \frac{1-\tau }{1-F(t)}\right)^{-\gamma}-1}{\gamma}.
\end{equation}
Under stationarity, the extreme $\tau$-quantile is also seen as $T$-return level associated to a certain return period $T$, where $\tau=1-1/T$. 
The return period is expressed on the scale of the original data that can be days, months or years. For example the $T$-return level is the value that is expected to be exceeded on average every $T$ years.

The domain of attraction condition \eqref{eq:EVTtheorem} is well-suited for continuous variables but can be  inadequate when applied to discrete variables \citep[e.g.][]{shimura2012discretization}. 
A useful approach between mathematical foundation and practical applicability is provided by the framework introduced by \citet{hitz_davis_samorodnitsky_2024} \citep[see also][]{chavez2022dGP,koh2023gradient} which we adopt and briefly summarize next {\color{black}{(see Section A of the supplement for details)}}.
Let $Z$ be a discrete random variable defined on $\{ 0,1,2,\dots\}$ and assume that its distribution is in the domain of attraction of the discrete-GP (d-GP) distribution, with $\gamma \geq 0$.
Specifically, let $\overline{Y}$ be a continuous random variable whose distribution $\overline{F}$ satisfies $\overline{F} \in \mathcal{D}(H_{\gamma})$, with $\gamma\geq 0$. Then, set $Z \overset{d}{=}\lfloor \overline{Y}\rfloor$, meaning that $\mathbb{P}(Z \geq z)=\mathbb{P}(\overline{Y} \geq z)$, in other terms the random variable $Z$ and its continuous extension $\overline{Y}$ are equal in distribution for $z \in \{ 0,1,2,\dots\}$, where $\lfloor x\rfloor$ is largest integer smaller or equal than $x$.
Because $\overline{F} \in \mathcal{D}(H_{\gamma})$, then we have that $Z$ is in the discrete domain of attraction, in symbols $Z \in \text{d-}\mathcal{D}(H_{\gamma})$.
Indeed, for a high threshold $t$ close to right-end point $y_E$ we have that $\mathbb{P}\left( \overline{Y}-t \geq y \mid \overline{Y} > t\right)$ $\approx \overline{H}_{\bftheta}\left(y\right)$, for $y>0$, where $\overline{H}_{\bftheta}(\cdot)=1-{H}_{\bftheta}(\cdot)$, and this result can be used to approximate the distribution of a discrete excess variable. Specifically, for a large threshold $t$ we have,
\begin{align}\label{eq:dgpcdf}
\mathbb{P}(Z-t \leq z \mid Z\geq t) &= \sum_{q=0}^{z} \mathbb{P}(\overline{Y}-t\geq q \mid \overline{Y} \geq t) - \sum_{q=0}^{z} \mathbb{P}(\overline{Y}-t\geq q+1 \mid\overline{Y} \geq t)\nonumber  \\
& \approx \sum_{q=0}^{z} \overline{H}_{\bftheta}\left(q \right)-\sum_{q=0}^{z} \overline{H}_{\bftheta}\left(q +1\right) \nonumber\\
& = G_{\bftheta}\left(z \right) := 1-\left(1+ \gamma \frac{z+1}{\sigma} \right)^{-\frac{1}{\gamma}}.
\end{align}
This result implies that the probability mass function of the d-GP distribution is $g_{\bftheta}(z)=\overline{H}_{\bftheta}(z) - \overline{H}_{\bftheta}(z+1)$.
Similarly to the continuous case, for a large enough value $z>t$ and $t$ approaching to $y_E$, then $\mathbb{P}(Z\leq z \mid Z\geq t)\approx G_{\bftheta}\left(z-t \right)$ and the tail of the unknown unconditional distribution function of $Z$ can be approximated as,
\begin{equation}\label{eq:discre:quant}
F(z) \approx F(t) + (1-F(t)){G}_{\bftheta} \left(z-t \right).
\end{equation}
Finally, for a quantile level $\tau$ close to one, we have that an approximation of the extreme $\tau$-quantile is given by
\begin{equation}\label{eq:quantextdisc}
Q(\tau) \approx t + \Bigg\lfloor \sigma \frac{\left( \frac{1-\tau }{1-F(t)}\right)^{-\gamma}-1}{\gamma}  \Bigg \rfloor- 1.
\end{equation}

%
%
\subsection{Nonstationary case}\label{sec:modnonstat}
%
%

The modelling in Section \ref{sec:homos} is suitable for stationary extremes over time. It can be extended
to the case of time-varying non-stationary extremes according to the heteroscedastic extremes framework in \citet{einmahl2016statistics}. In the application of Section \ref{sec:results}, daily losses and fatalities are not recorded at regular time intervals (i.e., every day), and the exact timing of each loss is unknown a priori. It is natural then to model the occurrence of losses and fatalities as a random process. In this context, the framework of heteroscedastic extremes extends to a more general setting through the {\it proportional tail model} introduced by \citet{dombry2023asymptotic}, which we briefly summarize next.

Consider the pair ($Y, \bfX)$, where $Y$ is a response variable whose distribution $F$ satisfies $F \in \mathcal{D}(H_{\gamma})$, with $\gamma\in\Real$, whose right-end point is $y_E$, $\bfX$ is $d$-dimensional random vector of covariates which we assume in $[0,1]^d$ without loss of generality and whose marginal law is denoted by $P(\diff x)$. We assume that the conditional distribution $F_{\bfx}(y)=\Prob(Y\leq y \mid \bfX=\bfx)$, shares the same right-end point and satisfies the \textit{tail proportionality} condition,
\begin{equation}\label{eq:scedasis}
\underset{y \to y_E}{\lim}\frac{1-F_{\bfx}(y)}{1- F(y)} = c(\bfx), \qquad \bfx \in [0,1]^d,
\end{equation}
where, $c(\bfx)$ is a bounded function, known as the {\it scedasis} function \citep{einmahl2016statistics}, which describes how the frequency of extreme events varies with $\bfx$. This condition implies that the tail of the conditional distribution $F_{\bfx}$ scales with $c(\bfx)$, while the tail heaviness, characterized by the extreme value index $\gamma$, remains unchanged across $\bfx$.

By applying the tail approximation in \eqref{eq:quant} to the present nonstationary case, together with \eqref{eq:scedasis}, we can approximate for a large threshold $t$ approaching to $y_E$ and $y>t$ and any arbitrary measurable set $B\subset [0,1]^d$ the tail probability $\Prob(Y>y, \bfX \in B)$ by $P^{\star}(B)(1-F(t))(1-H_{\bftheta}(y-t))$, where $P^\star(\diff \bfx)=c(\bfx)P(\diff \bfx)$ is the law of the \textit{concomitant covariates}, i.e.  the conditional law of $\bfX$ given that $Y>t$. This entails, first, that for $t \to y_E$ and $y>t$
\begin{equation}\label{eq:asym_ind}
\Prob(Y\leq y, \bfX\in B \mid Y>t)\approx P^\star(B)H_{\bftheta}(y-t),
\end{equation}
second, we have $\Prob(Y\leq y \mid \bfX \in B)\approx 1- P^{\star}(B)(1-F(t))(1-H_{\bftheta}(y-t))/P(B)$ known as the conditional proportional tail model \citep[see][for details]{dombry2023asymptotic}. When $B\downarrow \bfx$ we have $P^\star(B)/P(B)\approx c(\bfx)$ and therefore in this case the conditional proportional tail model approximation entails for $t\to y_E$ and $y>t$
\begin{equation}\label{eq:condtail}
\Prob(Y\leq y \mid \bfX = \bfx) \approx 1-(1-F(t)) c(\bfx) \left( 1 + \gamma\frac{y-t}{\sigma}  \right)_{+}^{-\frac{1}{\gamma}}.
\end{equation}
Accordingly, the $\tau$-quantile of the conditional distribution of $\mathbb{P}(Y \leq y \mid \bfX = \bfx)$ can be approximated for $\tau$ close to $1$ by
\begin{equation}
Q_{\bfx}(\tau)\approx t + \sigma\frac{\left( \frac{1-\tau}{1-F(t)} \frac{1}{c(\bfx)} \right)^{-\gamma} -1}{\gamma}.
\end{equation}
Furthermore, \cite{dombry2023asymptotic} proposed a statistical test to assess whether the concomitant covariates have a significant impact on the extremes of the response variable. If the covariates have no effect, then $c(\bfx)=1$ for all $\bfx\in[0,1]^d$. Accordingly, the hypothesis test is formulated as follows:
\begin{equation}\label{eq:test}
H_0: F_{P^\star} = F_P \quad \quad \text{versus} \quad \quad H_1: F_{P^\star} \neq F_P,
\end{equation}
where $F_{P^\star}$ and $F_P$ are the cumulative distribution functions of the laws $P^\star$ and $P$. \cite{dombry2023asymptotic} provided a Kolmogorov-Smirnov type of test for testing the validity of the null hypothesis $H_0$, which we summarize  in Section D of the supplement.

Finally, this nonstationary setting can be extended to the discrete case through the following simple argument.  Let $\overline{Y}$ be a random variable whose distribution $\overline{F}$ satisfies $\overline{F} \in \mathcal{D}(H_{\gamma})$, with $\gamma\geq 0$ and right-end point $y_E$. Assume that the proportional tail framework holds here with the response variable $\overline{Y}$. We recall that in Section \ref{sec:homos} we assume that $\Prob(\overline{Y}\geq z)=\Prob(Z \geq z)$ for $z \in \{0,1,2,\dots \}$ and analogously here we assume that $\Prob(\overline{Y}\geq z\mid \bfX=\bfx)=\Prob(Z \geq z\mid \bfX=\bfx)$ for $z \in \{0,1,2,\dots \}$. As a result the relationship \eqref{eq:scedasis} holds except for a finite set of points and  {the} conditional proportional tail model in the discrete case becomes for $t\to y_E$ and $y>t$
\begin{align}\label{eq:tailregdiscrete}
\Prob(Z \leq z \mid \bfX=\bfx) &\approx
1- (1-F(t)) c(\bfx) \left(1+\gamma\frac{z-t+1}{\sigma} \right)^{-\frac{1}{\gamma}}
\end{align}
and the extreme quantile corresponding to such a conditional distribution can be approximated for $\tau$ close to $1$ as
\begin{equation}\label{eq_cond_quantile}
Q_{\bfx}(\tau)\approx t +\Bigg\lfloor  \sigma\frac{\left( \frac{1-\tau}{1-F(t)} \frac{1}{c(\bfx)} \right)^{-\gamma} -1}{\gamma} \Bigg \rfloor -1.
\end{equation}

%
%
\section{Inference and prediction}\label{sec:mod:stat:inf}
%

%
%
\subsection{Stationary case}\label{sec_inference_stationary}
%

Let $\bfY_n=(Y_1,\dots,Y_n)$ be a sample of iid continuous random variables, whose unknown distribution $F$ satisfies $F\in \mathcal{D}(H_{\gamma})$. 
%
%
According to Section \ref{sec:homos}, the peak variables (peaks), extracted from the sample, approximately follow the GP distribution given in \eqref{eq:EVTtheorem}, provided a sufficiently high threshold is chosen, and they can be used to estimate the model parameter $\bftheta$. A common approach for threshold selection is to use a high quantile, i.e. $t=Q(\tau_I)$, where $\tau_I=1-k/n$ is an intermediate level. Specifically, we select the the number $k=k_n$ of peaks that we want to work with, named the {\it effective sample size}, according to the criterion  $k\to \infty$ as $n\to \infty$, while $k$ remaining small relative to the sample size $n$, i.e., $k=o(n)$.
As long as the {\it effective sample proportion} $\tau_I\cdot 100\%$ is strictly less than 100\%, then $Q(\tau_I)$ remains large but is still expected to lie within the observed sample on average. We estimate $Q(\tau_I)$ using the $(n-k)$th order statistic $Y_{n-k,n}$, where $Y_{1,n}\leq \cdots \leq Y_{n,n}$ denotes the $n$th ordered sample. Consequently, the top $k$ order statistics, $(Y_{n-i+1,n})_{1 \leq i\leq k}$, are used for parameter estimation.
A popular choice for $s(t)$ is $s(t)=a(F^{\leftarrow}(1-1/t))$, where $F^{\leftarrow}(\tau)=\inf(x:F(x)\geq \tau)$ is the left continuous inverse of $F$ and $a(\cdot)$ is the classical positive scaling function used in the EVT \citep[][Theorem 1.1.6]{haan2006extreme}. Estimating $F$ by the empirical distribution $F_n$ allows $s(t)\approx a(F_n^{\leftarrow}(1-1/Y_{n-k,n}))=a(n/k)$. In the sequel, the scale parameter $\sigma$ is representative of $a(n/k)$. Similar reasoning applies for selecting $t$ and $s(t)$ in the discrete case, where in this case the threshold is estimated using the $(n-k)$th ordered statistic $Z_{n-k,n}$, from the sample $\bfZ_n=(Z_{1}, \dots, Z_{n})$ of discrete variables.

We propose an empirical Bayes procedure for the estimation of $\bftheta$ and $Q(\tau_E)$, where $\tau_E$ is an extreme level, defined as $\tau_E=1-c/n$, with $c$ being a positive constant. 
A key benefit of the Bayesian approach is that it naturally enables the derivation of a predictive distribution for forecasting future tail events. We briefly overview the method and refer to \cite{dombry2023asymptotic} for a more detailed description, including a thorough investigation of its asymptotic theory and performance.
Firstly, the likelihood function for $\bftheta\in\Theta=\mathbb{R}\times(0,\infty)$ pertaining to the GP and d-GP models, based on the $k$ top order statistics, is defined as
\begin{equation*}
{L}_n(\bftheta)=
\begin{cases}
\prod_{i=1}^k h_{\bftheta}(Y_{n-k+i,n}-Y_{n-k,n}),\quad   & \text{in the continuous case},\\
\prod_{i=1}^k g_{\bftheta}(Z_{n-k+i,n}-Z_{n-k,n}), \quad & \text{in the discrete case},
\end{cases}
\end{equation*}
where $h_{\bftheta}(y-t)$ and $g_{\bftheta}(z-t)$ are the densities of GP and d-GP distributions, respectively. The Maximum likelihood (ML) estimator $\widehat{\bftheta}_n=(\widehat{\gamma}_n, \widehat{\sigma}_n)$ of $\bftheta$ is obtained by
%
\begin{equation}\label{eq:MLE}
\widehat{\bftheta}_n \in \underset{\bftheta \in \Theta}{\arg \max} \ {L}_n(\bftheta). 
\end{equation}
We also recall  if one desires to appeal to the asymptotic normality property of the ML estimator, then the parameter space needs to be restricted as $\Theta=(-1/2) \times (0,\infty)$ (e.g., \citealp{haan2006extreme} Ch 3, \citealp{dombry2023asymptotic}).
%

Secondly, we specify a prior distribution as follows. While $\gamma$ is a standard parameter that lies within the interior of the real line and is independent of the sample, the scale parameter $\sigma$ behaves differently. Since $\sigma$ represents $a(n/k)$, its value inherently depends on both $n$ and $k$. Consequently, the prior distribution for $\sigma$ must be constructed to account for these dependencies. Although specifying such a prior is a nontrivial task, \citet{dombry2023asymptotic, padoan2024empirical} propose a simple yet effective approach inspired by the empirical Bayes literature. In particular, the prior distribution on $\bftheta$ is obtained by defining its density as
\begin{equation}\label{eq:prior}
\pi(\bftheta) = \pi_{\text{sh}}(\gamma) \pi_{\text{sc}}^{(n)}(\sigma),
\end{equation}
for $n=1,2,\ldots$. In particular, $\pi_{\text{sh}}(\gamma)$ is a prior density of $\gamma$. Possible options for it are: informative proper prior density (e.g.,  normal, Student-$t$, Stable, etc.), noninformative improper prior density (e.g., uniform, Jeffreys, maximal data information, etc.). Instead, $ \pi_{\text{sc}}^{(n)}(\sigma)=\pi(\sigma/\widehat{\sigma}_n)/\widehat{\sigma}_n$, where $\pi(\sigma)$ is an informative proper prior density (e.g., gamma, Pareto, Weibull, etc.) and $\widehat{\sigma}_n$ is a consistent estimator of $\sigma$. In this way, distinct samples of different sample size $n$ lead to different values of $a(n/k)$, but nevertheless 
our prior distribution updates accordingly by means of $\widehat{\sigma}_n$, placing a bit of mass in a neighborhood of true value of $\sigma$, implying the consistency of the resulting posterior distribution \citep{dombry2023asymptotic}. Although the prior in \eqref{eq:prior} assumes independence between $\gamma$ and $\sigma$, \cite{dombry2023asymptotic, padoan2024empirical} demonstrated that this assumption does not have a substantial impact, as theoretical guarantees of accuracy remain valid, and, from a practical perspective, the results are highly accurate.

For all measurable sets $B\subseteq \Theta$ , the posterior distribution of $\bftheta$ is then defined as 
\begin{equation}\label{eq:statpost}
\varPi_n(B) = \frac{\int_{B}{L}_n({\bftheta}) \pi(\bftheta)\text{d}\bftheta}{\int_{\bftheta} {L}_n({\bftheta}) \pi(\bftheta)\text{d}\bftheta}.
\end{equation}
There is no closed-form expression for the posterior distribution $\varPi_n(B)$, so inference relies on the ability to generate samples from it. In particular, \cite{padoan2024empirical} proposed a Markov Chain Monte Carlo (MCMC) scheme for sampling from the posterior distribution under the block maximum approach, which was later extended to the POT method by \cite{dombry2023asymptotic}. {\color{black}{Details on the MCMC algorithm are given in Section C of the supplement}}. If $\overline{\bftheta}$ follows the posterior distribution $\varPi_n$, formula 
\begin{equation}\label{eq:RLstatest}
\overline{Q}(\tau_E)= 
\begin{cases}
Y_{n-k,n}+ \overline{\sigma}\left(\frac{{\tau^{\star}}^{-\overline{\gamma}}-1}{\overline{\gamma}}\right), & \text{in the continuous case},\\
Z_{n-k,n}+ \Bigg\lfloor\overline{\sigma}\left(\frac{{\tau^{\star}}^{-\overline{\gamma}}-1}{\overline{\gamma}}\right)\Bigg\rfloor-1, & \text{in the discrete case},
\end{cases}
\end{equation}
induces an approximative posterior distribution, say $\overline{\varPi}_n$, for the extreme quantile $Q(\tau_E)$. 
In particular, $\tau^\star=(1-\tau_E)/(1-\tau_I)$  is a parameter in $[0,1]$ that determines the severity of such a quantile. The closer $\tau^\star$ is to 0, the more extreme the quantile; conversely, as $\tau^\star$ approaches 1, the quantile becomes an intermediate quantile and so less extreme. \cite{dombry2023asymptotic} demonstrated that the posterior distribution $\overline{\varPi}_n$  implying a reliable inference of $Q(\tau_E)$ as it becomes increasingly accurate as the sample size $n$ grows. Given a sample $\overline{\bftheta}_i$, for  $i=1,\ldots,M$ with $M\geq 1$, has been generated from the posterior distribution $\varPi_n(\bftheta)$ a sample from  $\overline{\varPi}_n(\overline{Q}(\tau_E))$ is readily obtained applying \eqref{eq:RLstatest}.

%
We address now the problem of the prediction of future tail events.
Let $Y_{n+1}$ be an out-of-sample random variable representative future values that have not yet been observed, independent of the past sample $\bfY_n$. Since our focus is on extremes, we work with the predictive distribution of tail events $F^\star_{\tau_E}(y):=\Prob(Y_{n+1}\leq y\mid Y_{n+1}> t_E, \bfY_n)$. Here, $t_E$ is an extreme threshold, and we allow for either an intermediate or a proper extreme threshold, $t_E\geq t_I$, where the severity of future extremes is controlled by $\tau^\star$. Leveraging the threshold stability property \citep[][Ch. 1, 5]{falk2010}, the true unknown $F^\star_{\tau_E}$ can still be approximated by a GP distribution, but in this case, the scaling function is given by $s(t_E)=s(t_I)+\gamma(t_E-t_I)$, where $t_I$ is an intermediate threshold. The GP-based approximation of $F_{\tau_E}^\star$ and its density $f_{\tau_E}^\star$ are
\begin{equation*} \label{eq_predictive}
H_{\bftheta}^{\tau^\star}(y)=H_\bftheta\left(r_\bftheta (y-Y_{n-k,n};\tau^\star)\right),\quad
h_{\bftheta}^{\tau^\star}(y)=h_\bftheta\left(r_\bftheta (y-Y_{n-k,n};\tau^\star)\right){\tau^\star}^\gamma,
\end{equation*}
where
\begin{equation}\label{eq:stand_point}
r_\bftheta (y-Y_{n-k,n};\tau^\star)=(y-Y_{n-k,n}){\tau^\star}^\gamma-\sigma\frac{1-{\tau^\star}^\gamma}{\gamma}.
\end{equation}
%
%
Under the Bayesian approach the posterior predictive distribution and density
%
%
\begin{equation} \label{eq:post_pred_dist}
\widehat{F}_n^\star(y) = \int_{\Theta} H_{\bftheta}^{\tau^\star}(y)\varPi_n(\diff\bftheta),\quad
\widehat{f}_n^\star(y) = \int_{\Theta} h_{\bftheta}^{\tau^\star}(y)\varPi_n(\diff\bftheta)
%
\end{equation}
%
are estimators of  $F^\star_{\tau_E}$ and $f_{\tau_E}^\star$, respectively. \cite{dombry2023asymptotic} demonstrated that $\widehat{F}_n^\star$ is a consistent estimator $F^\star_{\tau_E}$ in Wasserstein distance, while \cite{padoan2023statistical} showed that $\widehat{f}_n^\star$ converges to the true predictive density in Hellinger distance, and, that extreme regions derived from $\widehat{f}_n^\star$, such as $(1-\alpha)\cdot100\%$ predictive intervals, for any $\alpha\in(0,1)$, are highly accurate for the large samples.
Furthermore, the posterior predictive distribution can also be used to compute point forecasts for the RL via the formula $(\widehat{F}_n^\star)^{\leftarrow}(1-\tau^\star)$, where again $\tau^\star=(1-\tau_E)/(1-\tau_I)$, but in this case, $\tau_E=1-1/T$, with return period $T=1,2,\ldots$ expressed in the data scale (e.g., days, months, or years). 

Similar reasoning applies in the discrete case, therefore we only shortly list the difference. The past sample and the out-of-sample variable are denoted by $\bfZ_n$ and $Z_{n+1}$, respectively, and the predictive distribution and mass function by $G^\star_{\tau_E}$ and $g^\star_{\tau_E}$. The threshold stability property remains valid, however, the corresponding predictive distribution is approximated by a d-GP distribution with a scaling function given by $s(t_E)=s(t_I)+\gamma(t_E-t_I+1)$, {\color{black}{see Section B of the supplement for details}}. According to \eqref{eq:dgpcdf}, the d-GP-based approximation of $G^\star_{\tau_E}$ and $g^\star_{\tau_E}$ are given by
$$
G_{\bftheta}^{{\tau}^\star}(z)= G_\bftheta\left(\Bigg\lfloor r_\bftheta (z-Z_{n-k,n};\tau^\star)\Bigg\rfloor\right),\quad g_{\bftheta}^{{\tau}^\star}(z)= g_{\bftheta}(\lfloor r_\bftheta (z-Z_{n-k,n};\tau^\star)\rfloor),
$$
where $r_\bftheta$ is defined in \eqref{eq:stand_point} and $G_{\bftheta}^{{\tau}^\star}$ and $g_{\bftheta}^{{\tau}^\star}$ are defined in Section \ref{sec:homos}. Estimators of true predictive distribution and mass function are given by the posterior predictive distribution and mass function, which are obtained by replacing   in \eqref{eq:post_pred_dist}  $H_{\bftheta}^{\tau^\star}$ and $h_{\bftheta}^{\tau^\star}$ by $G_{\bftheta}^{{\tau}^\star}$ and $g_{\bftheta}^{{\tau}^\star}$, respectively.
%
In practice, given a sample $\overline{\bftheta}_i$ with $i=1,\ldots,M$ from the posterior distribution $\varPi_n(\bftheta)$, the posterior predictive density and mass functions can be approximated by computing the Monte Carlo approximations
\begin{equation}\label{eq:approx_post_pred}
\widehat{f}_n^\star(y)\approx \frac{1}{M}\sum_{i=1}^M h_{\overline{\bftheta}_i}^{\tau^\star}(y),\quad
\widehat{g}_n^\star(y)\approx \frac{1}{M}\sum_{i=1}^M g_{\overline{\bftheta}_i}^{\tau^\star}(y).
\end{equation}
%

%
%
\subsection{Nonstationary case}\label{sec_inference_nonstationary}
%

Let $(Y_1,\bfX_1),\ldots, (Y_n,\bfX_n)$ be a sample, where $Y_i$ follows an unknown distribution $F$, satisfying $F\in \mathcal{D}(H_{\gamma})$, $\bfX_i$ follows an unknown law $P(\diff x)$ and the conditional distribution $F_{\bfx_i}(y)=\Prob(Y_i\leq y \mid \bfX_i=\bfx_i)$ satisfies the tail proportionality condition in \eqref{eq:scedasis}, for each $i=1,\ldots, n$. For the inference on $(\bftheta, c(\bfx))$ we use a Bayesian approach that relies on the pairs of excess variables and concomitant covariates $(Y_{n-i+1,n}-Y_{n-k,n},\bfX_{n-i+1,n})_{1\leq i \leq k}$, where $(\bfX_{n-i+1,n})_{1\leq i \leq k}$ are the elements of the sequence $\bfX_{1,n},\ldots,\bfX_{n,n}$ that are associated to the top $k$ variables $Y_{n-i+1,n}$, $1\leq i \leq k$. The law of the concomitant variables is denoted by $P^\star$. 
To simplify the inferential procedure, it is convenient to specify a prior distribution of the form $\varLambda(\diff \bftheta, \diff P^\star)=\varPi(\diff\bftheta)\varPhi(\diff P^\star)$, where $\varPi$  is the prior on $\bftheta$ defined through the formula in \eqref{eq:prior}, and $\varPhi$ is a prior on $P^\star$, with $P^\star$ lying in the space of of Borel probability measures on $[0,1]^d$, say $\mathscr{P}$. A natural choice for  $\varPhi$ is $\varPhi(\diff P^\star)=\text{DP}(\diff P^\star,\rho)$, where DP denotes the Dirichlet process and $\rho$ is a finite positive measure on measurable sets $B\subset [0,1]^d$\citep[see, e.g.,][Ch. 4.1]{ghosal2017}. According to this prior and the asymptotic independence model in \eqref{eq:asym_ind}, it follows that the posterior distribution on $(\bftheta, P^\star)$ is for all measurable sets $(B\times C)\subset \Theta\times\mathscr{P}$,
\begin{equation*}\label{eq:posterior_pot_cov}
	\varLambda_n(B\times C)=\varPi_n(B)\varPhi_n(C).
\end{equation*}
This also splits into a product between the posterior distribution on $\bftheta$ and the posterior distribution on $P^\star$, where the latter becomes by conjugacy of the Dirichlet process (see Ch 4.6 in \citealp{ghosal2017}) equal to  $\varPhi_n(C)=\text{DP}(C;\rho+k\Prob_n^\star)$, with parameter $\rho+k\Prob_n^\star$, where $\Prob_n^*(\cdot)=k^{-1}\sum_{i=1}^k\indic(\bfX_{n-i+1,n}\in\cdot)$ is the empirical measure associated to the concomitant covariates. Although $\varPi_n$ and $\varPhi_n$ are dependent random measures, \cite{dombry2023asymptotic} demonstrated that handling the two posterior distributions separately is asymptotically justified and it works very well in practice. Inference on $\bftheta$ has been already described at the beginning of this section. Here, we describe the inference on $P^\star$ and more importantly on $c(\bfx)$. Let  $B_n$ be a sequence of sets containing $\bfx$ and with a decreasing volume. A possible choice is the ball $B_n=B(\bfx, r_n)=\{\bfy\in[0,1]^d:\|\bfy-\bfx\|\leq r_n\}$ with center $\bfx$ and deterministic radius $r_n$, so that the ball volume is the same for all $\bfx\in[0,1]^d$.
Since $c(\bfx)\approx P^\star(B_n)/P(B_n)$ for $n \to\infty$, then if we knew $P$, then the Dirichlet Process prior on $P^\star$ would induce a prior on $P^\star(B_n)/P(B_n)$ and inference on $c(\bfx)$ could be done by the resulting conjugate posterior. Because $P$ is unknown in practice, we first estimate it using $\widehat{p}_n(\bfx) =n^{-1}\sum_{i=1}^{n}\indic(\bfX_i\in B_n)$ and then specify a data-dependent prior on $c(\bfx)$, for a given $\bfx$, by leveraging on the formula $c(\bfx):=P^\star(B_n)/\widehat{p}_n(\bfx)$, and then deriving its posterior distribution $\varPsi_n$. Despite the use of a data-dependent prior, \cite{dombry2023asymptotic} demonstrated that the resulting posterior distribution retains desirable accuracy properties for large samples. Since $\varPhi_n$ is  a Dirichlet process and $\varPsi_n$ depends on it, sampling from the Dirichlet-multinomial distribution and the computation of its density and other related quantities
can be done using the {\tt R} package {\tt extraDistr} \cite{Tymoteusz20}.

Let $\overline{\bftheta}$ and $\overline{c}(\bfx)$, for a given $\bfx$, be distributed according to the posterior distributions $\varPhi_n$ and $\varPsi_n$, respectively, then by \eqref{eq:condtail} and \eqref{eq_cond_quantile} we have that the formula
\begin{equation}\label{eq:extreme_quantile_conditional}
\overline{Q}_{\bfx}(\tau_E) = 
\begin{cases}
Y_{n-k,n} + \overline{\sigma}\frac{\left( \frac{\tau^\star}{\overline{c}(\bfx)} \right)^{-\overline{\gamma}} -1}{\overline{\gamma}} & \text{in the continuous case},\\
Z_{n-k,n} + \Bigg\lfloor \overline{\sigma}\frac{\left(\frac{\tau^\star}{\overline{c}(\bfx)} \right)^{-\overline{\gamma}} -1}{\overline{\gamma}}\Bigg\rfloor - 1 & \text{in the discrete case},
\end{cases}
\end{equation}
induces an approximate posterior distribution for the extreme conditional quantile $Q_\bfx(\tau_E)$, in the continuous and discrete case, which can be used for its estimation.
%
We complete this section describing the prediction task in the nonstationary case. Given a past sample $(\bfY_n, \bfX^{(n)})$, where $ \bfX^{(n)}=(\bfX_1,\ldots,\bfX_n)$, we consider an independent out-of-sample response variable and covariate vector $(Y_{n+1}, \bfX_{n+1})$, representative of future values not yet observed. We focus on the conditional distribution $F^\star_{\tau_E}(y\mid\bfx)=\Prob(Y_{n+1}\leq y \mid Y_{n+1}>Q_{\bfx}(\tau_E), \bfX_{n+1}=\bfx, \bfX_n, \bfY_n)$, for all $y>Q_{\bfx}(\tau_E)$ and $\bfx\in(0,1)^d$. We call it the predictive distribution of tail events conditionally to covariates. The GP-based approximation of it and its density functions are
\begin{equation*} \label{eq_predictive_nonstat}
H_{\bftheta}^{\tau^\star}(y\mid \bfx)=H_\bftheta\left(r_\bftheta\left(y-Y_{n-k,n}; \frac{\tau^\star}{c(\bfx)}\right)\right),\quad
h_{\bftheta}^{\tau^\star}(y\mid \bfx)=h_\bftheta\left(r_\bftheta\left(y-Y_{n-k,n}; \frac{\tau^\star}{c(\bfx)}\right)\right),
\end{equation*}
where $r_\bftheta$ is defined in \eqref{eq:stand_point}.
%
Bayesian estimators of 
$F^\star_{\tau_E}$ and $f^\star_{\tau_E}$  are given by the posterior predictive distribution and density functions conditionally to covariates
\begin{equation}\label{eq:pred_covariates}
\widehat{F}_n^{\star}(y\mid\bfx)={\int_{\Omega}}
H_{\bftheta}^{\tau^\star}(y\mid \bfx)
\Phi_n(\diff c(\bfx)){\Pi_n(}\diff \bftheta),\;
\widehat{f}_n^{\star}(y\mid\bfx)={\int_{\Omega}}
h_{\bftheta}^{\tau^\star}(y\mid \bfx)
\Phi_n(\diff c(\bfx)){\Pi_n(}\diff \bftheta),
\end{equation}
where $\Omega=\Theta\times\mathscr{P}$. Once again a similar argument also applies in the discrete case, where the past sample and out-of-sample response variable are denoted by $\bfZ_n$ and $Z_{n+1}$, and the corresponding predictive distribution and mass function by $G^\star_{\tau_E}$ and $g^\star_{\tau_E}$. The latter can be approximated using d-GP-based approximations by
\begin{equation*} \label{eq_predictive_discrete}
G_{\bftheta}^{\tau^\star}(y\mid \bfx)=G_\bftheta\left(\Bigg\lfloor r_\bftheta\left(z-Z_{n-k,n}; \frac{\tau^\star}{c(\bfx)}\right)\Bigg\rfloor\right),\;
g_{\bftheta}^{\tau^\star}(z\mid \bfx)=g_\bftheta\left(\Bigg\lfloor r_\bftheta\left(z-Z_{n-k,n}; \frac{\tau^\star}{c(\bfx)}\right)\Bigg\rfloor\right),
\end{equation*}
where $G_{\bftheta}$ and $g_{\bftheta}$ are defined in Section \ref{sec:homos}.
Estimators of $G^\star_{\tau_E}$ and $g^\star_{\tau_E}$, denoted by $\widehat{G}_n^{\star}$ and $\widehat{g}_n^{\star}$, are obtained by replacing  in \eqref{eq:pred_covariates} $H_{\bftheta}^{\tau^\star}$ and $h_{\bftheta}^{\tau^\star}$ by $G_{\bftheta}^{\tau^\star}$ and $g_{\bftheta}^{\tau^\star}$, respectively. 

Finally, for practical purposes,
given a sample $\overline{\bftheta}_1,\ldots,\overline{\bftheta}_M$ from $\varPi_n$ and $\overline{c}_1(\bfx),\ldots,\overline{c}_M(\bfx)$ from $\varPsi_n$, for all $\bfx\in[0,1]^d$,
estimators $\widehat{f}_n^{\star}$ and $\widehat{g}_n^{\star}$ in \eqref{eq:pred_covariates} of the true predictive density $f^\star_{\tau_E}$ and mass function $g^\star_{\tau_E}$, conditionally to covariates, can be approximated by the Monte Carlo approximations
\begin{equation}\label{eq:approx_pred_covariates}
\begin{aligned}
\widehat{f}^{\star}_n(y\mid \bfx)&\approx \frac{1}{M} \sum_{i=1}^M h_{\overline{\bftheta}_i}^{\tau^\star}\left(r_\bftheta\left(y-Y_{n-k,n}; \frac{\tau^\star}{\overline{c}_i(\bfx)}\right)\right),\\
\widehat{g}^{\star}_n(y\mid \bfx)&\approx \frac{1}{M} \sum_{i=1}^M g_{\overline{\bftheta}_i}^{\tau^\star}\left(\Bigg\lfloor r_\bftheta\left(z-Z_{n-k,n}; \frac{\tau^\star}{\overline{c}_i(\bfx)}\right)\Bigg\rfloor\right).
\end{aligned}
\end{equation}
%
%
%

%
%
\section{Analysis of extreme economic losses and fatalities}\label{sec:results}
%
%
We apply the methodology outlined in Sections \ref{sec:homos} and \ref{sec:modnonstat} to 
analyse extreme losses and fatalities recorded in the dataset described in Section \ref{sec:data}.
After removing missing values 2,341 daily observations for losses and 4,039 observations for fatalities remain. In this section we use the abbreviations Symmetric 95\% Confidence Intervals (S-95\%CI), Asymmetric 95\% Credible Intervals (A-95\%CI) and Asymmetric 95\% Predictive Intervals (A-95\%PI).

%
%
\subsection{Stationary case}\label{sec:res:stat}
\begin{figure}[t!]
\begin{center}
\includegraphics[width=0.32\textwidth, page=11]{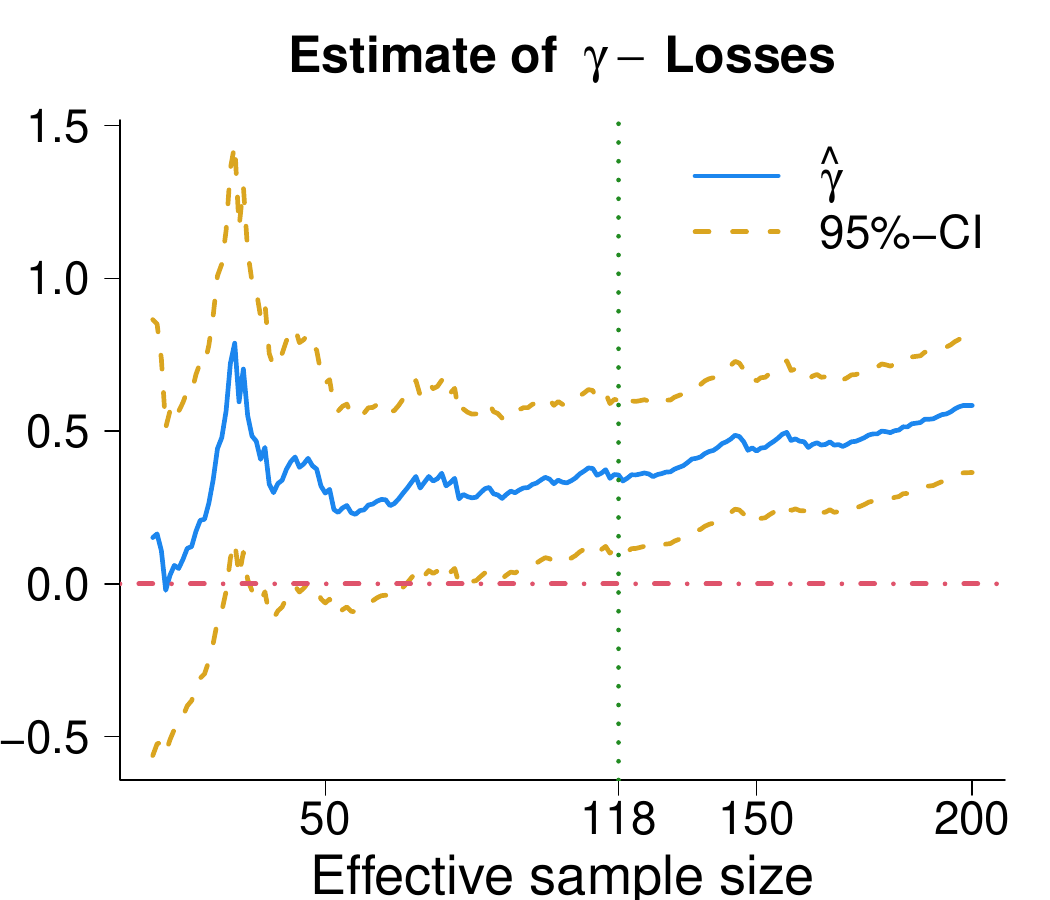}
\includegraphics[width=0.32\textwidth, page=12]{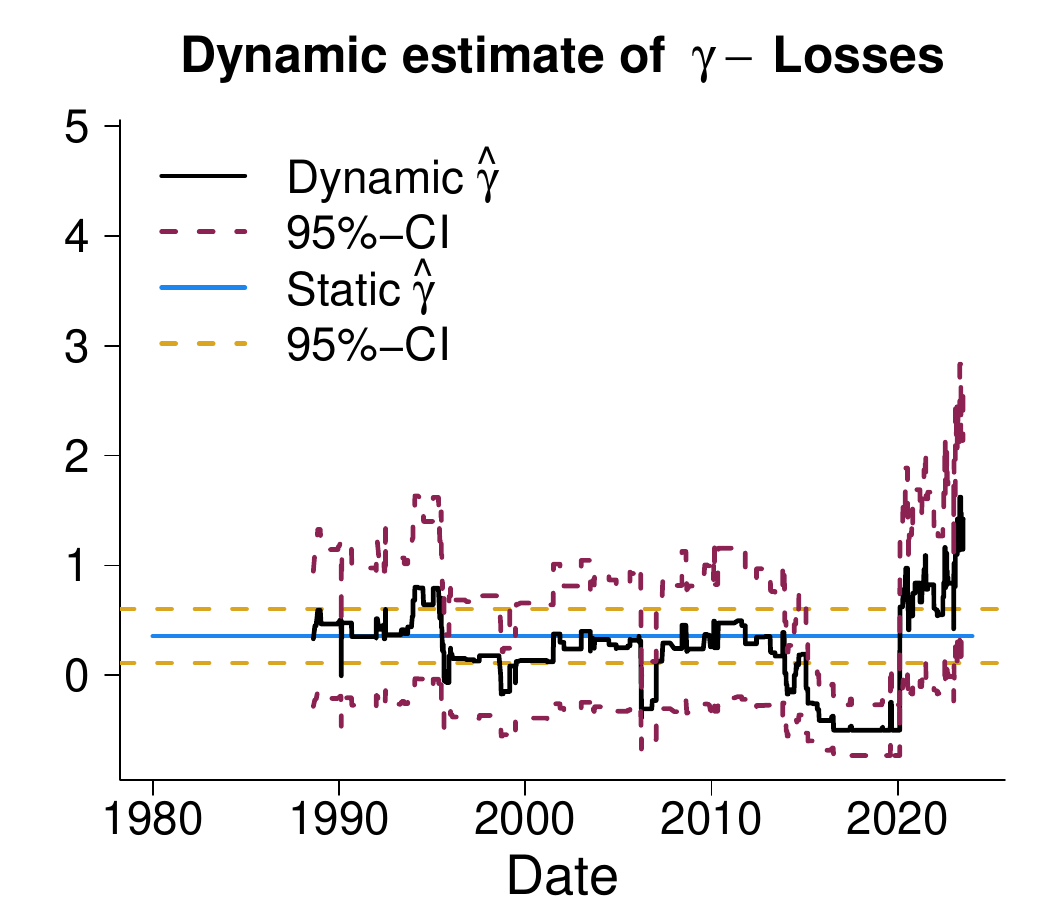}\\
\includegraphics[width=0.32\textwidth, page=13]{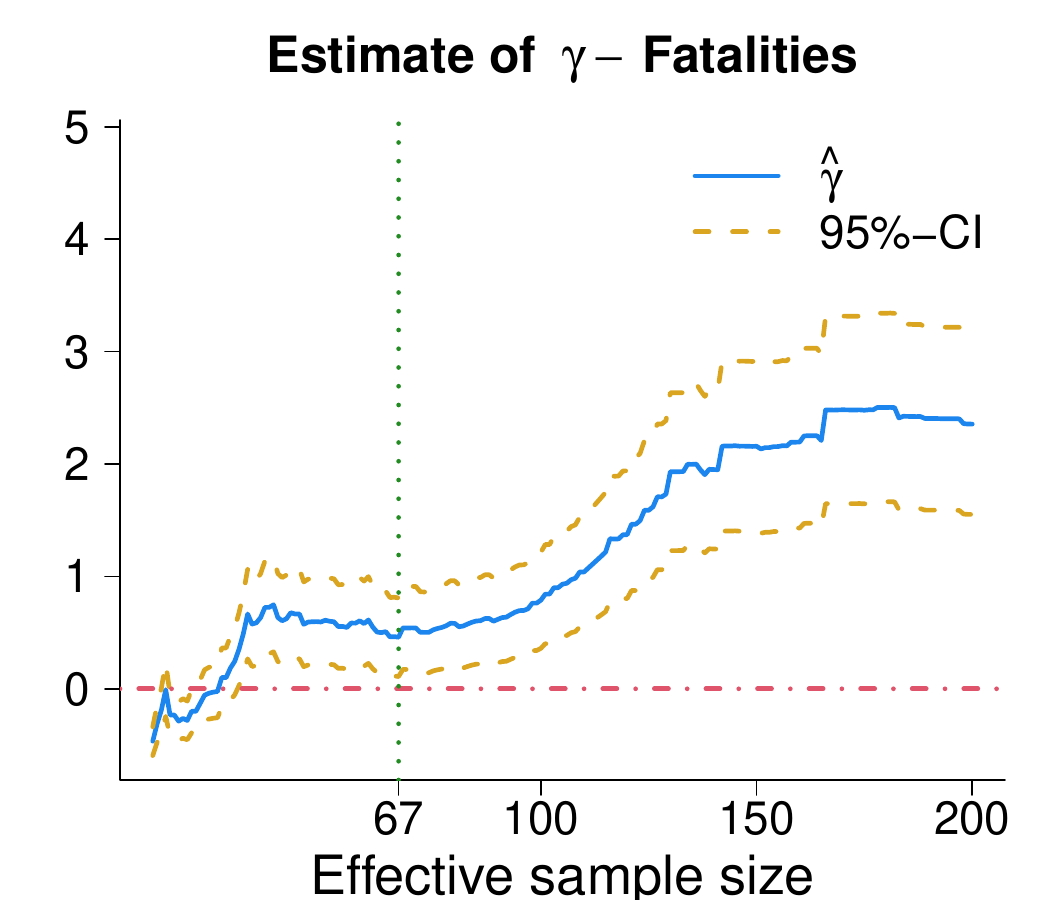}
\includegraphics[width=0.32\textwidth, page=14]{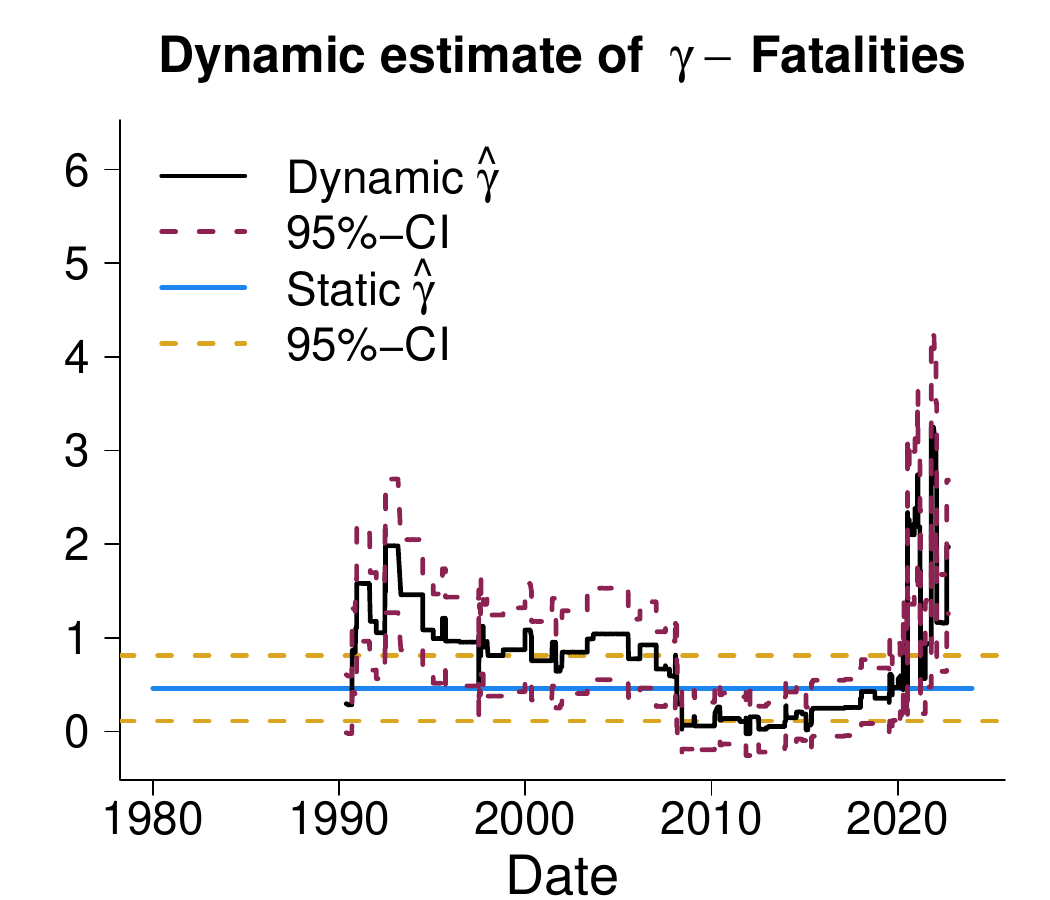}
\caption{Left panels show the ML estimates of $\gamma$ (blue solid line) and S-95\%CI (yellow dashed line) of losses (top row) and fatalities (bottom row) for different values of $k$. Right panels show the dynamic ML estimates of $\gamma$ obtained with a moving window (black solid line) for losses (top panel) and fatalities (bottom panel) and S-95\%CI (dark red dashed line).\label{fig:tailindic}}
\end{center}
\end{figure}

%
%
We start the analysis assuming for simplicity that losses and fatalities come from stationary distributions in the domain of attraction of the GP and d-GP models.
We first fit the GP and d-GP models to losses and the  fatalities, respectively, using the Maximum Likelihood (ML) approach.
The left column of Figure \ref{fig:tailindic} displays the estimates of $\gamma$ for $k$ varying from $10$ to $200$, for losses (top panel) and fatalities (bottom panel). For losses, we select the value $k = 118$ as it falls within a range where the estimates of $\gamma$ are stable (vertical green lines), and obtain with it an effective sample fraction $(1 - \tau_I) \cdot 100\% \approx 5\%$. 
The intermediate threshold is estimated as $y_{n-k,n}=$ \euro1.34 billion.
%
For fatalities, we select $k = 67$ which implies a sample fraction of $(1 - \tau_I) \cdot 100\% \approx 2\%$ and an estimate 
of the intermediate threshold of $z_{n-k,n}=418$ death.
With this setting, we obtain the estimates $\widehat{\gamma} =  {0.36}$ for losses and $\widehat{\gamma} =0.46$ for fatalities. Relying on 95\% confidence intervals (dashed yellow lines), we find empirical evidence supporting the heavy-tailed nature of both distributions.
We now want to assess whether the extreme value index $\gamma$ changes over time. A simple approach is illustrated in the right column of Figure \ref{fig:tailindic}, which displays dynamic estimates of $\gamma$ obtained using a moving window of approximately 800 observations, maintaining the same effective sample fractions as previously considered.
The dynamic estimates of $\gamma$ (black solid line) fluctuate around the static estimate (blue solid line), with both of their respective S-95\%CI—yellow for the static estimate and red dotted lines for the dynamic estimate—showing considerable overlap. These results provide no empirical evidence supporting a time-varying $\gamma$, and in this regard the stationary approach used in this section seems to be partially justified (see Section \ref{sec:res:nonstat} for an alternative non-stationarity assumption).

%
%
\begin{figure}[ht!]
\begin{center}
\includegraphics[width=0.32\textwidth, page=15]{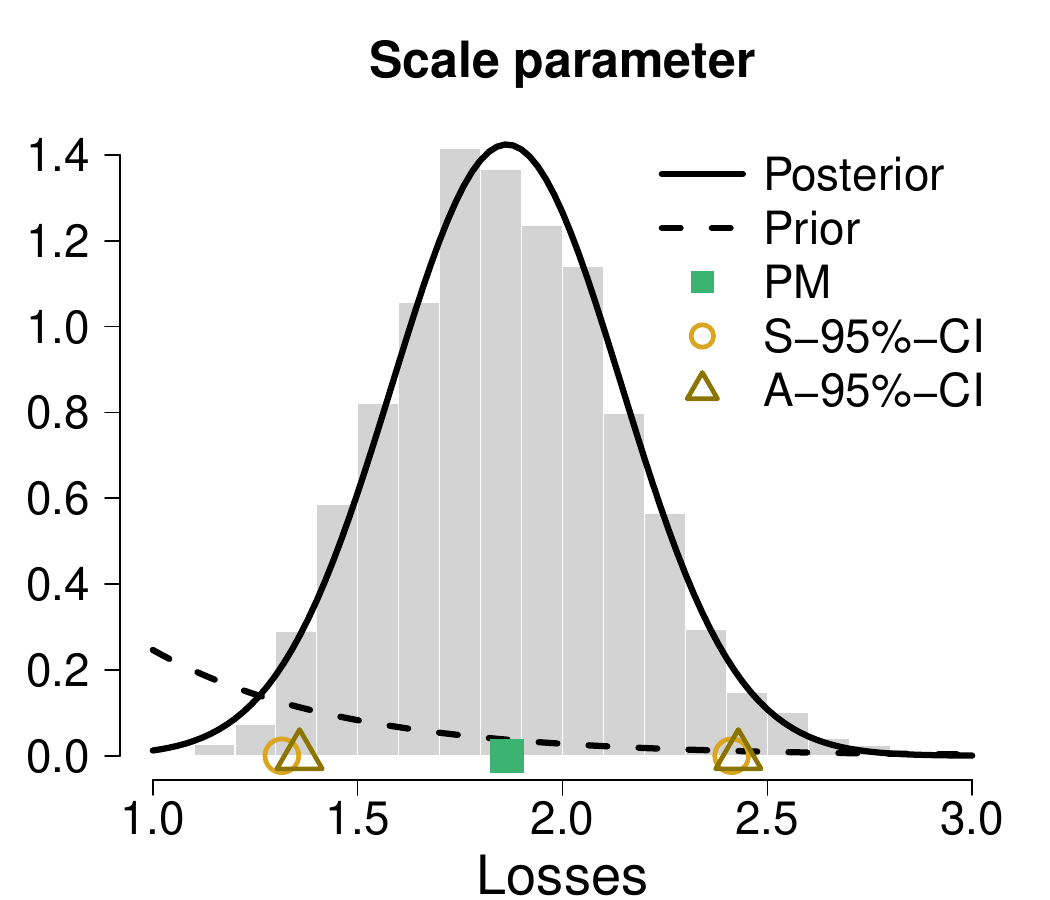}
\includegraphics[width=0.32\textwidth, page=16]{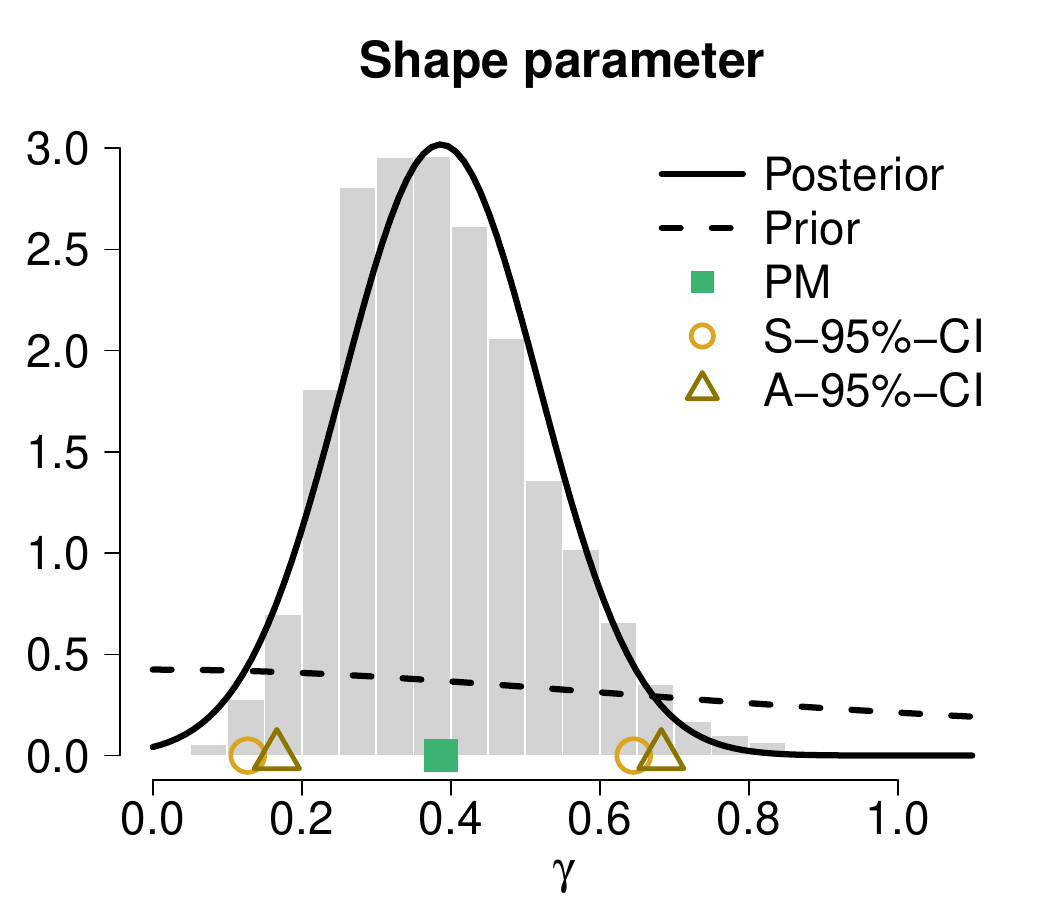}\\
\includegraphics[width=0.32\textwidth, page=20]{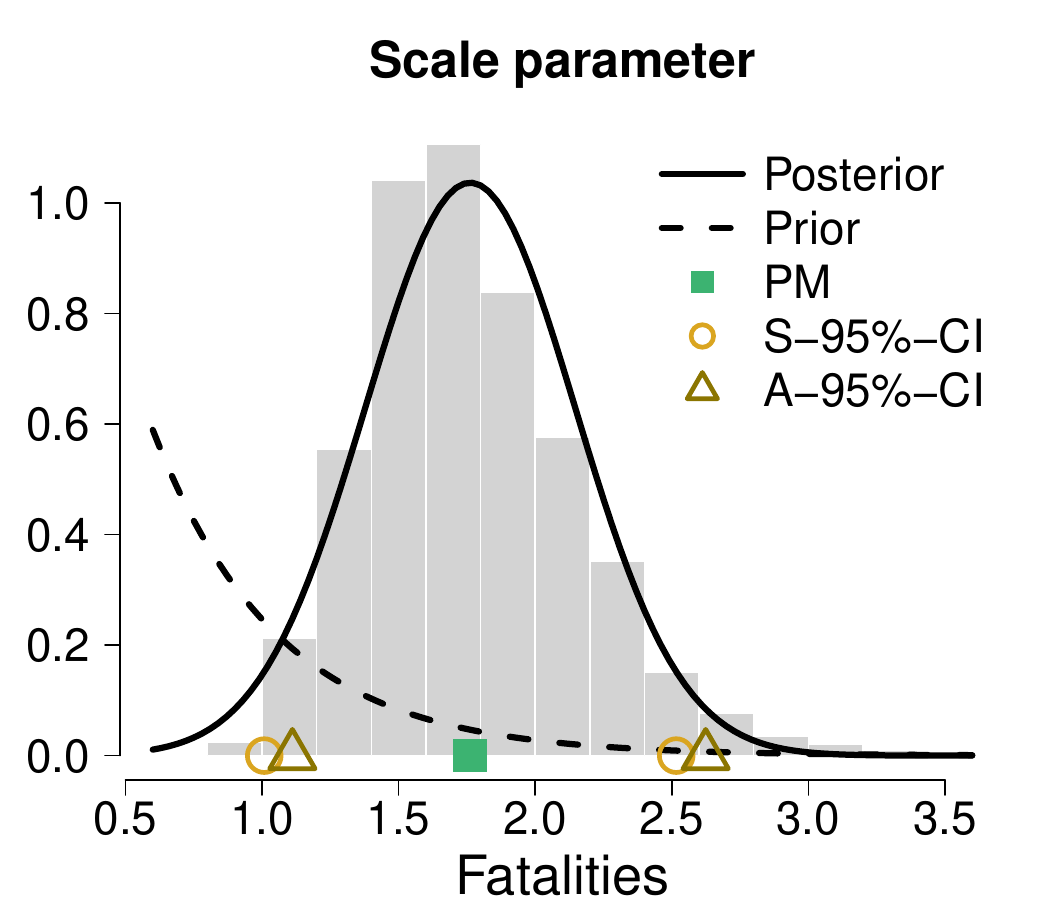}
\includegraphics[width=0.32\textwidth, page=21]{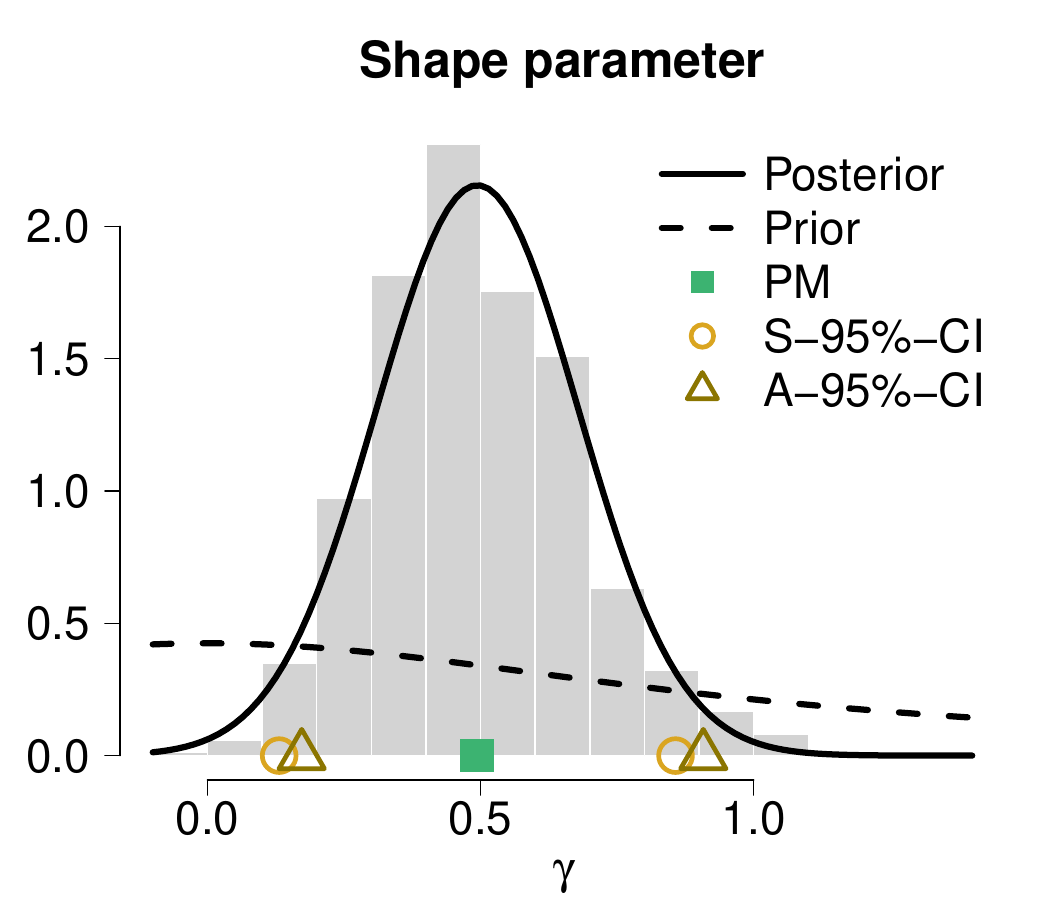}\\
\includegraphics[width=0.32\textwidth, page=18]{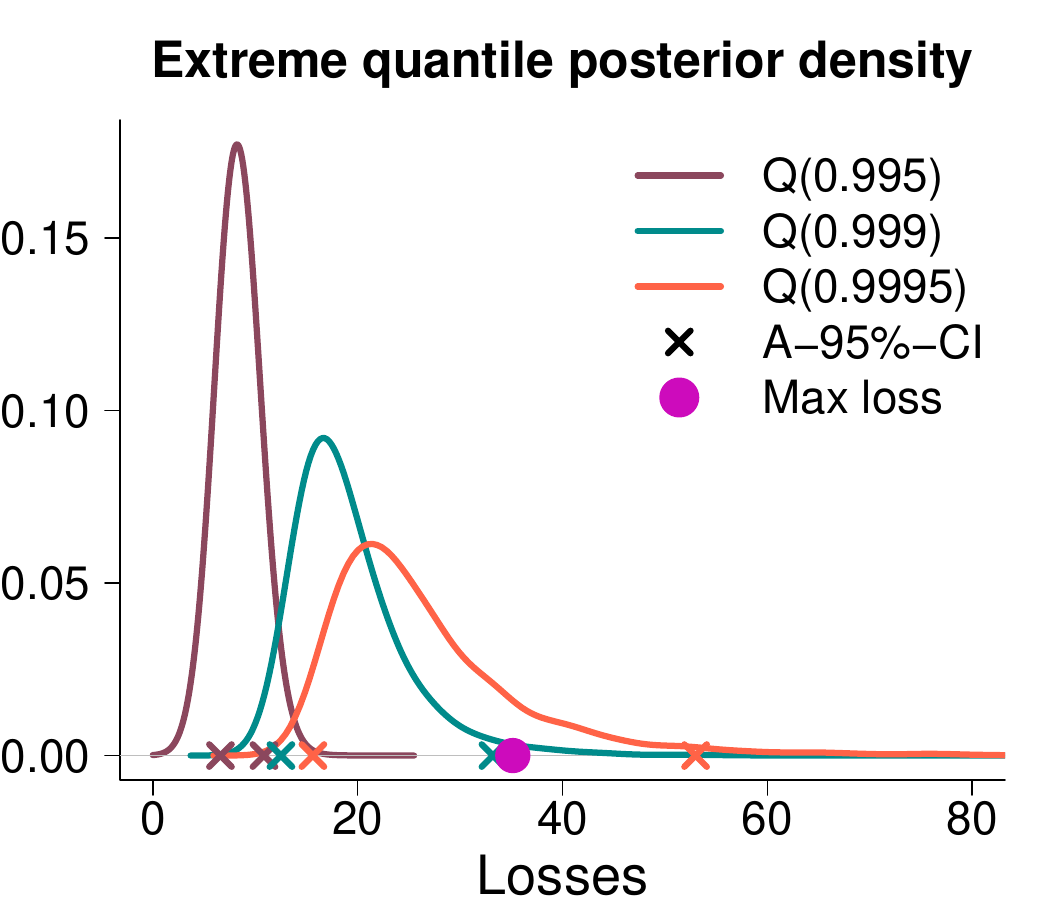}
\includegraphics[width=0.32\textwidth, page=23]{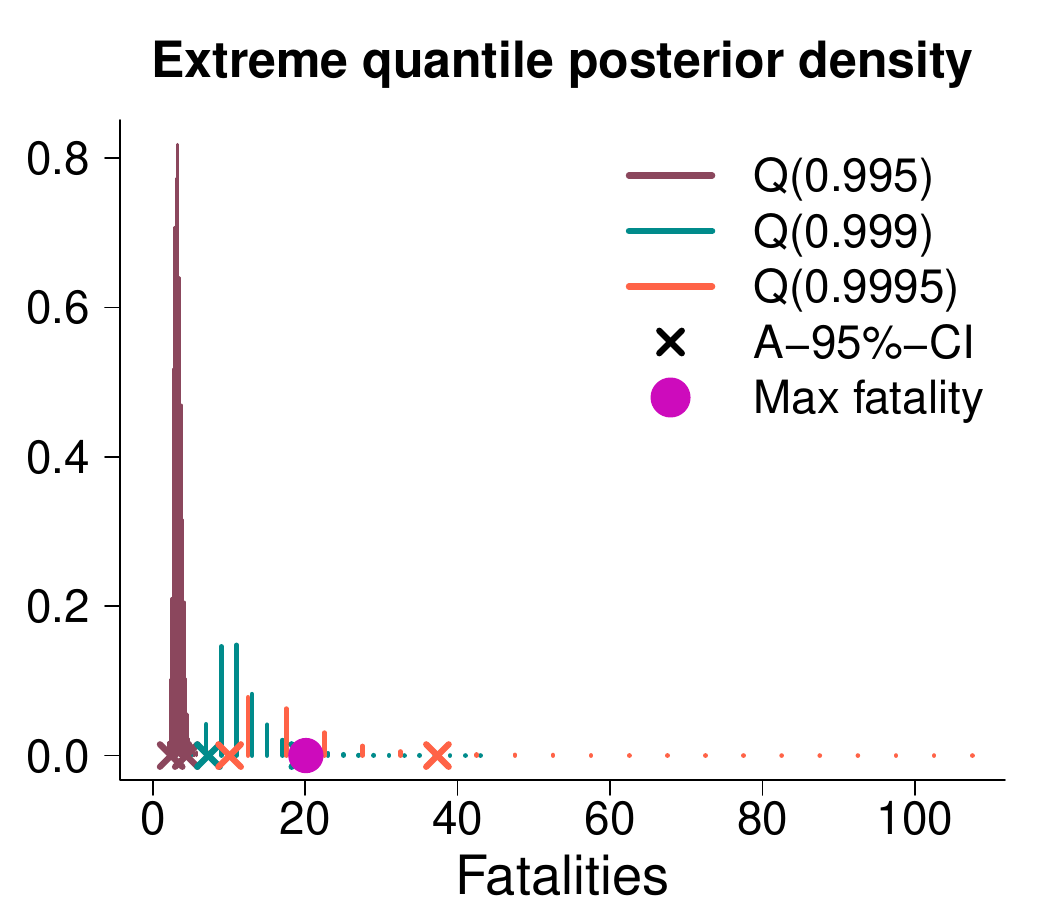}\\
\includegraphics[width=0.32\textwidth, page=17]{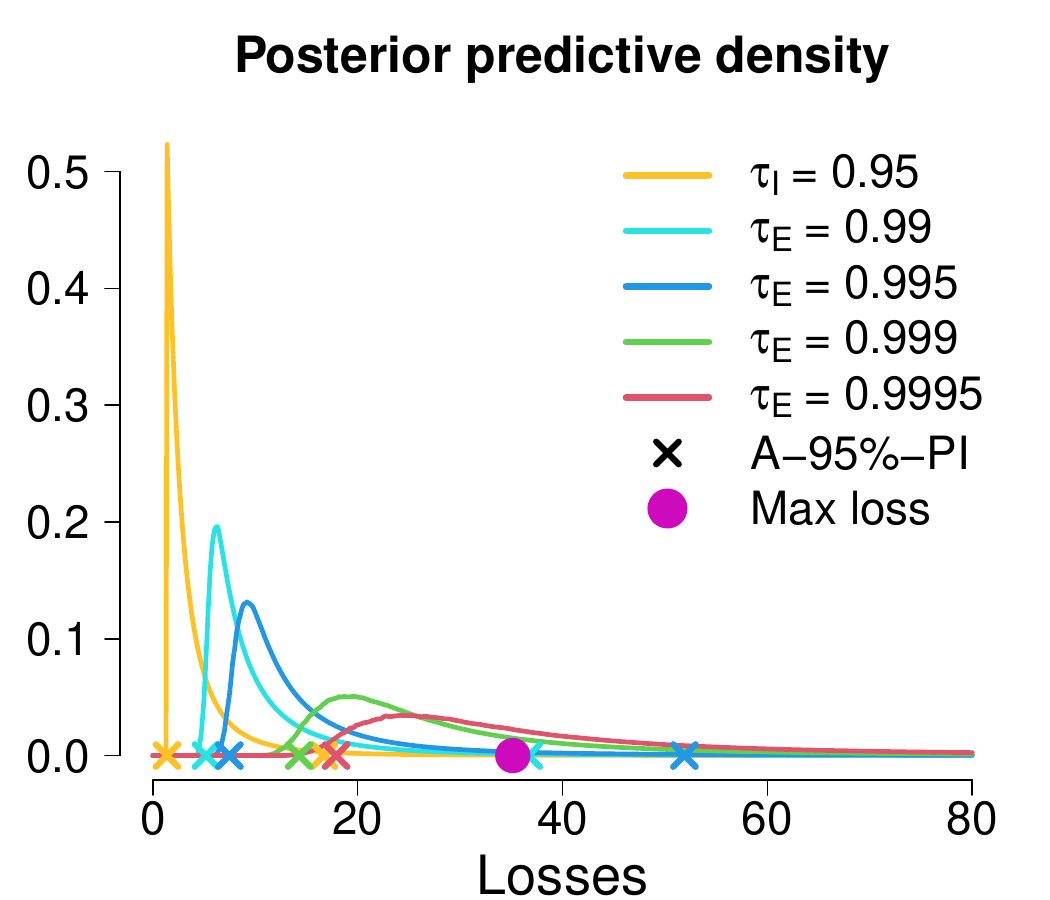}
\includegraphics[width=0.32\textwidth, page=22]{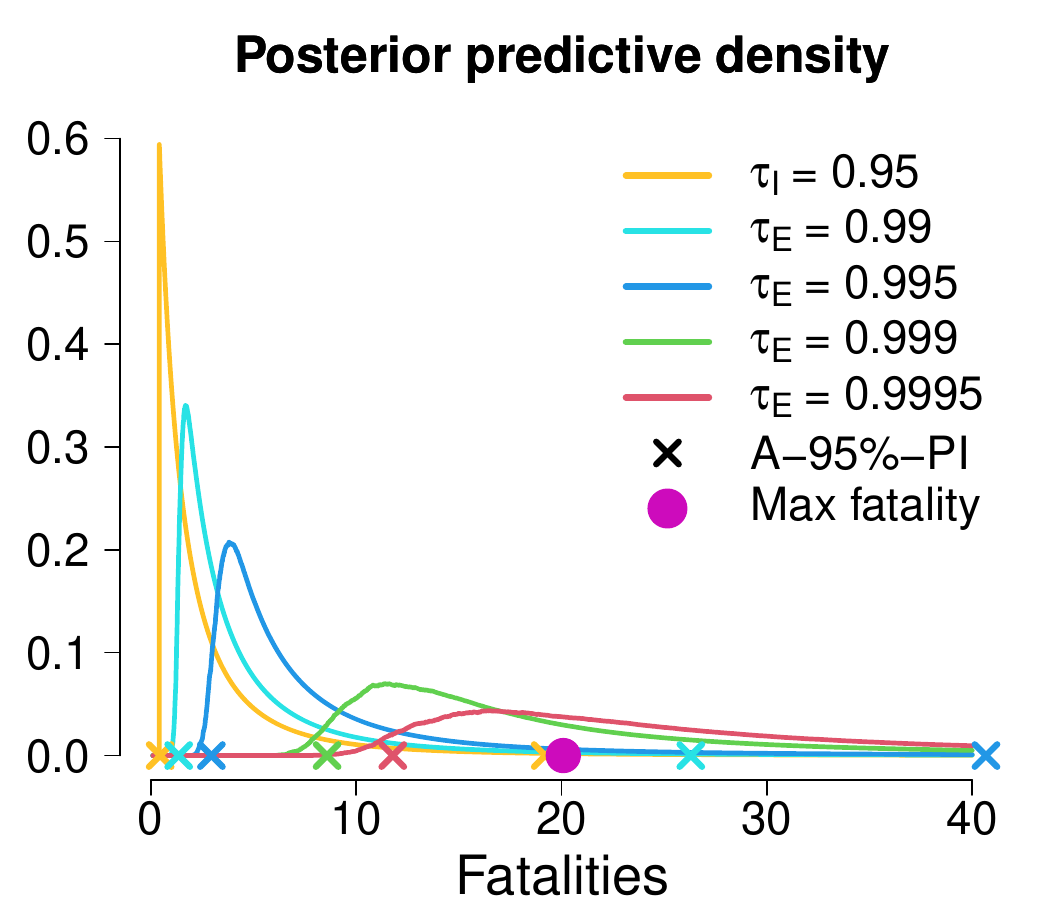}\\
\includegraphics[width=0.32\textwidth, page=19]{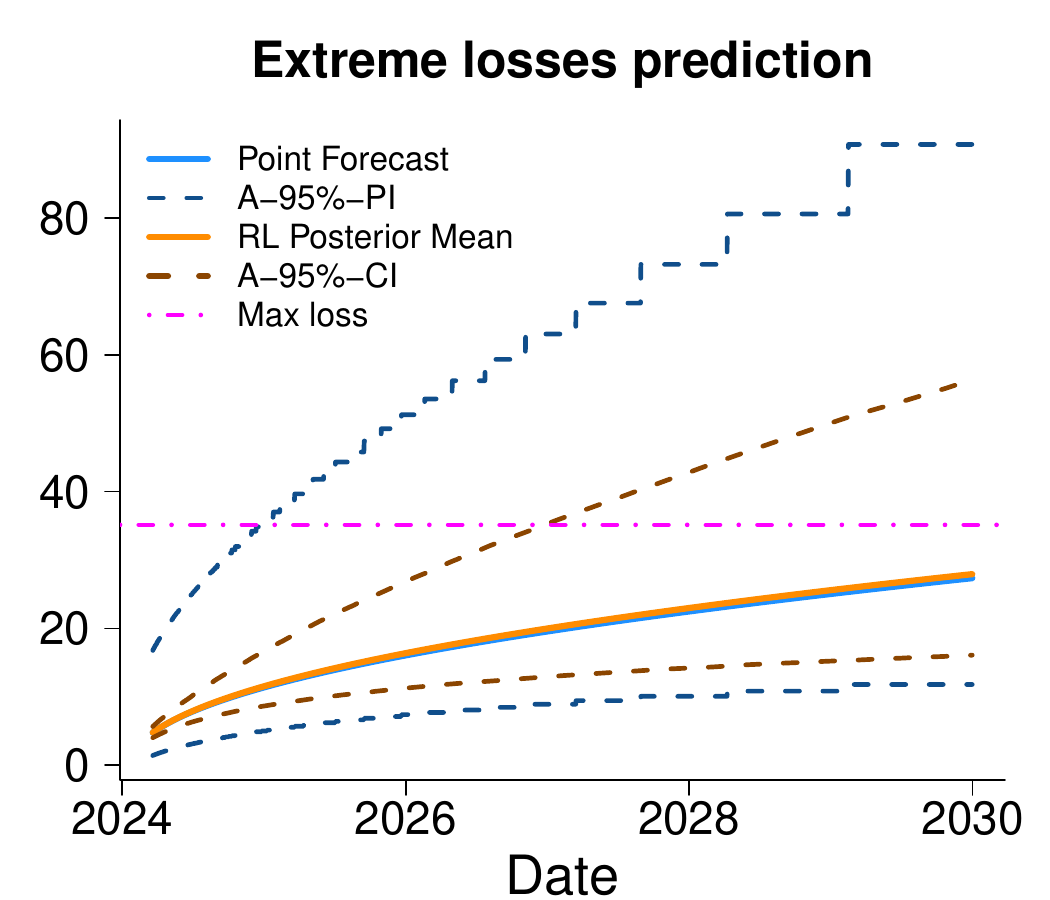}
\includegraphics[width=0.32\textwidth, page=25]{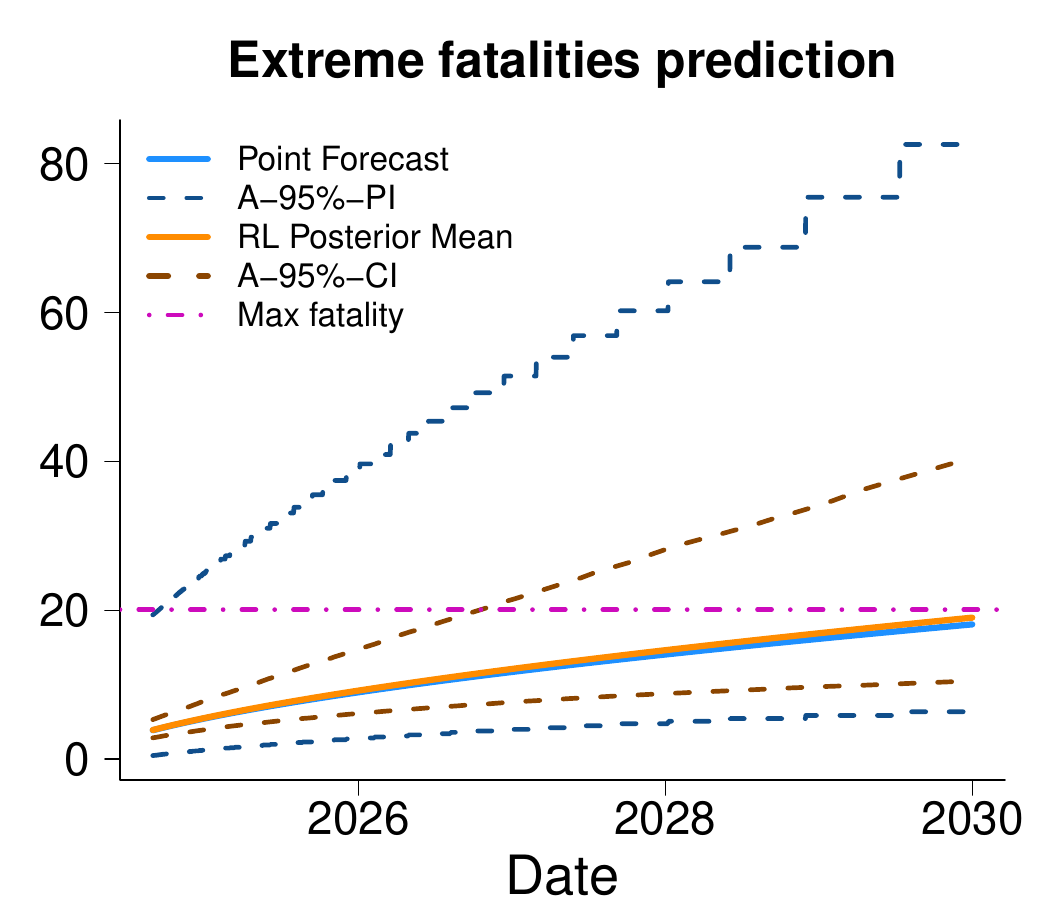}
\caption{Posterior (black solid line) and prior (black dashed line) densities of $\sigma$ (left panels) and $\gamma$ (right panels) parameters of the GP model (first and second raw for losses and fatalities, respectively). Posterior density of extreme quantiles (third row) for losses (left) and fatalities (right). Crosses depict the A-95\%CI. Posterior predictive density of future extreme losses (fourth row) for losses (left) and fatalities (right). Crosses depict the A-95\%PI. Posterior mean of Return Levels (RL) and RL point forecast (fifth row) and corresponding A-95\%CI and A-95\%PI.}
\label{fig:uncPoTest}
\end{center}
\end{figure}
%
%
We apply the Bayesian procedure outlined in Section \ref{sec:mod:stat:inf} to analyse losses and fatalities. The prior densities in equation \eqref{eq:prior} are specified as follows: $\pi_{\text{sh}}(\gamma)=(1-T_1(-1))^{-1}t_{1}\indic(-1 < \gamma < \infty)$, where $\indic$ is the indicator function, $T_{\nu}$ and $t_{\nu}$ are Student-t distribution and density functions, with $\nu$ degrees of freedom, respectively. For the scale parameter, we define: $\pi_{\text{sc}}\left({\sigma}\right)=\text{Ga}(\sigma; 1, \widehat{\sigma}_n)/\widehat{\sigma}_n$, where $\widehat{\sigma}_n$ is the ML estimator of $\sigma$, $\text{Ga}(\sigma; 1, \widehat{\sigma}_n)$ denotes a Gamma density with unit-shape and scale equal to $\widehat{\sigma}_n$.
The top-left and top-right panels of Figure \ref{fig:uncPoTest} report for the losses the empirical posterior densities of the scale and shape parameters, respectively, obtained simulating $M=20000$ particles $\overline{\bftheta}_i$, with $i=1,\ldots,M$, from the posterior distribution of GP model's parameters, by the MCMC algorithm \textcolor{black}{reported in Section C of the supplement}.
For $\sigma$, the posterior mean and A-95\%CI are \euro1.9 billion and [\euro1.36, \euro2.43] billion. For $\gamma$, the posterior mean and A-95\%CI are 0.4 and $[0.17, 0.68]$. The latter does not contain the value zero or negative values, supporting once more that the losses distribution can be considered heavy-tailed. Although the upper endpoint of the distribution is infinite, the mean remains finite. This emerges from the credible interval, which does not contain the value one. This feature is particularly useful as it ensures that the expected loss remains insurable by insurance companies. For the discrete case, the panels in the second row of Figure \ref{fig:uncPoTest} display the posterior density for $\sigma$ and $\gamma$, respectively. For $\sigma$, the posterior mean and A-95\%CI are  {1,762} deaths and $ {[1111, 2624]}$ deaths. For $\gamma$, the posterior mean and A-95\%CI are  {0.49} and $[0.17,  {0.91}]$. These results also confirm that the distribution of fatalities is heavy-tailed, and, although in this case the tail is heavier than that of losses, the expected number of deaths is still finite.

We then consider the intermediate level $\tau_I \cdot 100\% = 95\%$  {(losses) and 98\% (fatalities)} and the extreme levels $\tau_E \cdot 100\% = 99.5\%, 99.9\%, 99.95\%$, and by plugging into formula \eqref{eq:RLstatest} the simulated values $\overline{\bftheta}_i$, for $i=1,\ldots,M$, we obtain corresponding empirical posterior distributions for extreme quantiles. The third-left and right panels of Figure \ref{fig:uncPoTest}  display  in colored lines the resulting probability density (losses) and mass (fatalities) functions. The A-95\%CI (colored crosses) derived from them, support that the  largest historical recorded loss (fatality) of \euro35.15 billion (20,089 deaths), reported with the pink circle, is plausible only at the most extreme level, $\tau_E \cdot 100\%= 99.95\%$. This finding indicates that the largest historical loss (fatality) was indeed an exceptionally rare event, as it is expected to be exceeded in the future approximately with the tiny frequency of 0.5\%.  

Since the extrapolation of extreme quantiles may underestimate the uncertainty about the severity of future extreme events, we now analyse them through a proper predictive approach. 
We revisit the previously defined intermediate and extreme levels to specify suitable conditioning thresholds. The  upper (lower) section of Table \ref{tab:pfdist} summarizes the results for the losses (fatalities). In the intermediate case, we obtain a point forecast of \euro1.34 billion (418 deaths) and a A-95\%PI of approximately [\euro1.40; \euro17.00] billion. While these results are aligned with the extrapolation ones, in the extreme case different conclusion are achieved. We first recall that the posterior mean of extreme thresholds is \euro5.52, \euro8.32, \euro19.10 and \euro26.73 billion (1,426; 3,284; 11,398 and 17,929 deaths), respectively, while the A-95\%CI are displayed in the second column of the table. 
\begin{table}[t!]
\centering
\renewcommand{\arraystretch}{1.5} 
 \begin{tabular}{ccc}
 \toprule
  \multicolumn{3}{c}{Economic losses forecasts}\\
 \toprule 
 $\tau_I$&  $t_I$ & A-95\%-PI \\ 
 \midrule
 95.00\%  &   \euro\;1.34&  [\euro~1.39; \,\euro16.71] \\
 \cdashlinelr{1-3}
  $\tau_E$ & $t_E$-A-95\%-CI  & A-95\%-PI\\
 \cdashlinelr{1-3}
      \vspace{-0.2cm}
   99.00\% & [\euro~4.58; \, \euro~6.66]&  [\euro~5.20; \, \euro~36.70] \\ 
   \vspace{-0.2cm}
   99.50\%  &				[\euro~6.61; \, \euro10.86]				& [\euro~7.48; \,  \euro~51.90] \\
      \vspace{-0.2cm}
99.90\% & 					[\euro12.51; \, \euro33.21]				&[\euro14.29; \, \euro121.45] \\ 
\vspace{-0.2cm}
99.95\% & 				[\euro15.62; \, \euro52.99]				&[\euro17.90; \, \euro179.13]\\
\vspace{-0.5cm} &&\\
\bottomrule
\multicolumn{3}{c}{Fatalities forecasts}\\
 \toprule
  $\tau_I$& $t_I$ & A-95\%-PI \\ 
 \midrule
 98.00\%  &   418 ds&  [~~4 {59} ds; \,~19, {210} ds] \\
 \cdashlinelr{1-3}
  $\tau_E$& $t_E$-95\%-CI & A-95\%-PI\\
 \cdashlinelr{1-3}
      \vspace{-0.2cm}
   99.00\% & [1, 092 ds; \, ~1, 854 ds]&  [~1,362 ds; \, ~26,298 ds] \\ 
   \vspace{-0.2cm}
   99.50\%  &				[2,4{1}5 ds; \, ~4, {400} ds]				& [~ {2,954} ds; \,  ~ {40,670} ds] \\
      \vspace{-0.2cm}
99.90\% & 					[7, {311} ds; \,  {19,690} ds]				&[~8,5{88} ds; \,  {120,384} ds] \\ 
\vspace{-0.2cm}
99.95\% & 				[ {10,084} ds; \, 37,346 ds]				&[11,{786}; ds \, 1{97,084} ds]\\
\vspace{-0.5cm} &&\\
\bottomrule
  \end{tabular}
\caption{The upper section of the table presents the A-95\% PI in billion of euros (third column), computed by conditioning on a point estimate of the intermediate $t_I$ and extreme $t_E$ thresholds, whose posterior distribution is estimated and of which the A-95\%-CI is reported (second column). Intermediate $\tau_I$ and extreme $\tau_E$ levels are reported in the first column. The lower section of the table provides the results for the number of deaths (ds).}
\label{tab:pfdist}
\end{table}
%
%
%
The fourth-left and right panels of Figure \ref{fig:uncPoTest} show the posterior predictive density and mass function of future tail events for losses and fatalities, obtained using the Monte Carlo approximations in \eqref{eq:approx_post_pred}, provided that they exceed the aforementioned thresholds.
%
Accordingly, while the largest historical loss (fatality) of \euro35.15 billion (20,089 deaths) does not appear plausible when conditioning to the intermediate threshold, it does so when conditioning to all the extreme thresholds, see the A-95\%PI reported in third column of the Table \ref{tab:pfdist} for comparison. 
This analysis pushes risk severity to the far tail of the losses (fatalities) distribution, defining the worst-case scenario based on future losses (fatalities) exceeding predetermined thresholds with low frequencies of 1\%, \SI{5}{\permille}, \SI{1}{\permille}, and \SI{0.5}{\permille}, respectively. However, if these thresholds are breached, future losses (fatalities) will fall with 95\% probability within an interval of length \euro31.51, \euro44.42, \euro107.15, and \euro161.23 billion (24,936; 37,716; 111,796 and 185,298 deaths), respectively—capturing the inherent uncertainty of such an ambitious predictive goal.
The scenarios offer a valuable trade-off by highlighting both the rarity of certain future losses (fatalities) and the uncertainty surrounding their severity. 

We complete the analysis estimating the expected time horizon after which severe losses and fatalities are likely to occur. This is achieved by computing the {Return Level (RL) associated with the daily return period $T$. In this regard, we set $\tau_E=1-1/T$, where  $T \in \{\mathbb{N}_+: T>1/(1-\tau_I)\}$ and we consider then a time horizon spanning from March 20, 2024, to December 31, 2029,  {for the losses (corresponding to $T=80,\ldots, 365 + 1,\ldots, 365\times 5$), and from August 29, 2024, to December 31, 2029, for the fatalities (corresponding to $T=242,\ldots, 365 + 1,\ldots, 365\times 5$)}. The results are displayed in the bottom panel of Figure \ref{fig:uncPoTest}. The orange solid and brown dashed lines are the mean and the A-95\%CI obtained from the posterior distribution of RL, respectively. The   blue solid line shows the RL point forecast obtained as the ($1-\tau^\star$)-quantile of the posterior predictive distribution, estimated using the approximations in \eqref{eq:approx_post_pred}, where $\tau^\star=(1-\tau_E)/(1-\tau_I)$. 
In this case, to assess the uncertainty regarding the RL variations over time, we compute a A-95\%PI (dashed blue lines) based on the predictive distribution of tail events. The intermediate level is set as  $\tau_I=1-4/T$, ensuring that both $\tau_I$ and $\tau_E$ increase over time while maintaining the ratio $\tau^\star= 1/4$. This guarantees that the sample fraction $1 - \tau_I$ is four times the exceedance probability $1 - \tau_E$.

The  largest historical loss (fatality) seems to be an implausible event in the next six years according to Bayesian point estimates and forecasts of the RL. However, when accounting for estimation uncertainty, the A-95\%CI reveal that such an unprecedented loss (fatality) could still occur by late 2026. Instead, 
when accounting for estimation and aleatoric uncertainty, the A-95\%PI reveal an expected waiting time as early as the end of 2024.
Further details can be found in Table \ref{tab:PFstat}, which presents the specific values of the point estimate and forecast, along with A-95\%CI and A-95\%PI.

\begin{table}[t!]
\centering
\begin{tabular}{ccccc}
\toprule
\multicolumn{5}{c}{Economic losses forecasts}\\
\toprule
$T$ & RL & A-95\%-CI & RL & A-95\%-PI\\
\cdashlinelr{1-5}
Year & Posterior Mean && Point Forecast & \\
\midrule
2024 & \euro11.52    & [\euro~8.64; \euro16.64]    & \euro11.36   & [\euro~5.00; \euro35.53]\\
2025 & \euro16.37    & [\euro11.26; \euro26.84]   & \euro16.05   & [\euro~7.39; \euro51.25]   \\
2026 & \euro19.98    & [\euro12.92; \euro35.27]   & \euro19.55   & [\euro~8.91; \euro63.06]  \\
2027 & \euro22.97    & [\euro14.20; \euro42.81]   & \euro22.47   & [\euro10.06; \euro73.24]  \\
2028 & \euro25.59    & [\euro15.21; \euro50.03]   & \euro25.04   & [\euro10.83; \euro80.62]  \\
2029 & \euro27.93    & [\euro16.09; \euro56.40]   & \euro27.35   & [\euro11.80; \euro90.79] \\
\bottomrule
\multicolumn{5}{c}{Fatalities forecasts }\\
\toprule
$T$ & RL & A-95\%-CI & RL & A-95\%-PI\\
\cdashlinelr{1-5}
Year & Posterior Mean && Point Forecast & \\
\midrule
2024 & 5,5{40} ds    & [~3,9{62} ds; ~7, 843 ds]    & 5, {473} ds   &  [1, {199} ds; 24, {935} ds]\\
2025 & 9,2{07} ds   & [~6, {136} ds; 14, {767} ds]   & 8, {981} ds   &    [2,7{30} ds; 3{8,515} ds] \\
2026 & 12,1{15} ds   & [~7, {659} ds; 21, {425} ds]   & 11, {711} ds  & [4,0{00} ds;  {51,466} ds]\ \\
2027 & 14, {635} ds   & [~8, {803} ds; 2{8,150} ds]   & 1{4,055} ds  & [4, {768} ds;  {60,236} ds] \\
2028 & 16, {915} ds   & [~9, {736} ds; 3{4,076} ds]   & 16, {167} ds  & [5, {873} ds; 7 {5,498} ds] \\
2029 & 1{9,013} ds   & [10, {491} ds;  {40,500} ds]   & 1 {8,106} ds  & [6, {369} ds;  {82,583} ds]\\
\bottomrule
\end{tabular}
\caption{The upper section of the table presents the estimated RL based on extreme quantiles posterior mean (second column) along with the A-95\%CI (third column). It also includes the point forecast from the posterior predictive distribution (fourth column) and the associated the A-95\%PI (fifth column). Estimates are given for December 31st of each reported year (first column). The lower section of the table provides the estimates for fatalities.}
\label{tab:PFstat}
\end{table}

%
\subsection{Nonstationary case}\label{sec:res:nonstat}
%

We extend the analysis in Section \ref{sec:res:stat} by accounting for the time variation in the frequency of peaks. As shown in the bottom panels of Figure \ref{fig:bubbles}, the relative frequency of peak losses displays a declining trend over time, while the frequency of fatalities fluctuates throughout most of the period.
We incorporate the time-varying frequency of peaks by employing the proportional tail model introduced in Section \ref{sec:modnonstat}, where we assume that the conditional distribution $F_x$ depends on a random covariate $X \in [0,1]$, representing the time coordinate, and then apply the empirical Bayes procedure described in Section \ref{sec_inference_nonstationary} to fit the model to the observed peak losses and fatalities. We work with a random time, because losses (fatalities) do not occur every day but at unknown at priory times.
\begin{table}[ht!]
\centering
\begin{tabular}{rlll}
\toprule
\multicolumn{1}{c}{} & \multicolumn{1}{c}{Test statistic} & \multicolumn{1}{c}{Critical value} & \multicolumn{1}{r}{$p$-value} \\ \midrule  
Losses & 1.874                               & 1.227                      & 0.001 \\ 
Fatalities & 1.433                     & 1.286                     & 0.018 \\ \bottomrule     
\end{tabular}
\caption{Hypothesis testing for the assumption of homoscedasticity in extreme losses (top row) and fatalities (bottom row) over time (null hypothesis) versus the heteroscedasticity assumption. The observed test statistic (second column) is compared to the critical value (third column), while the corresponding $p$-value is reported in the fourth column.
\label{tab:test}}
\end{table}
%
%
\begin{figure}[h!]
\begin{center}
\includegraphics[width=0.32\textwidth, page=26]{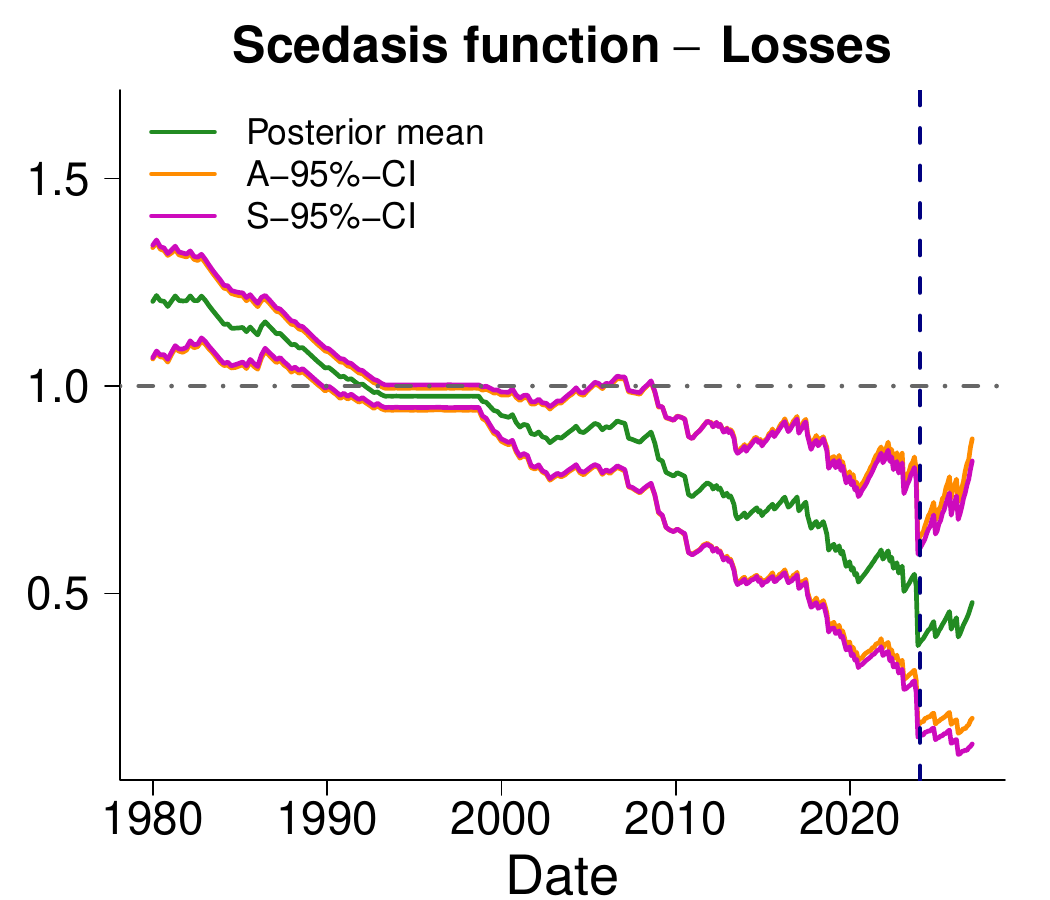}
\includegraphics[width=0.32\textwidth, page=28]{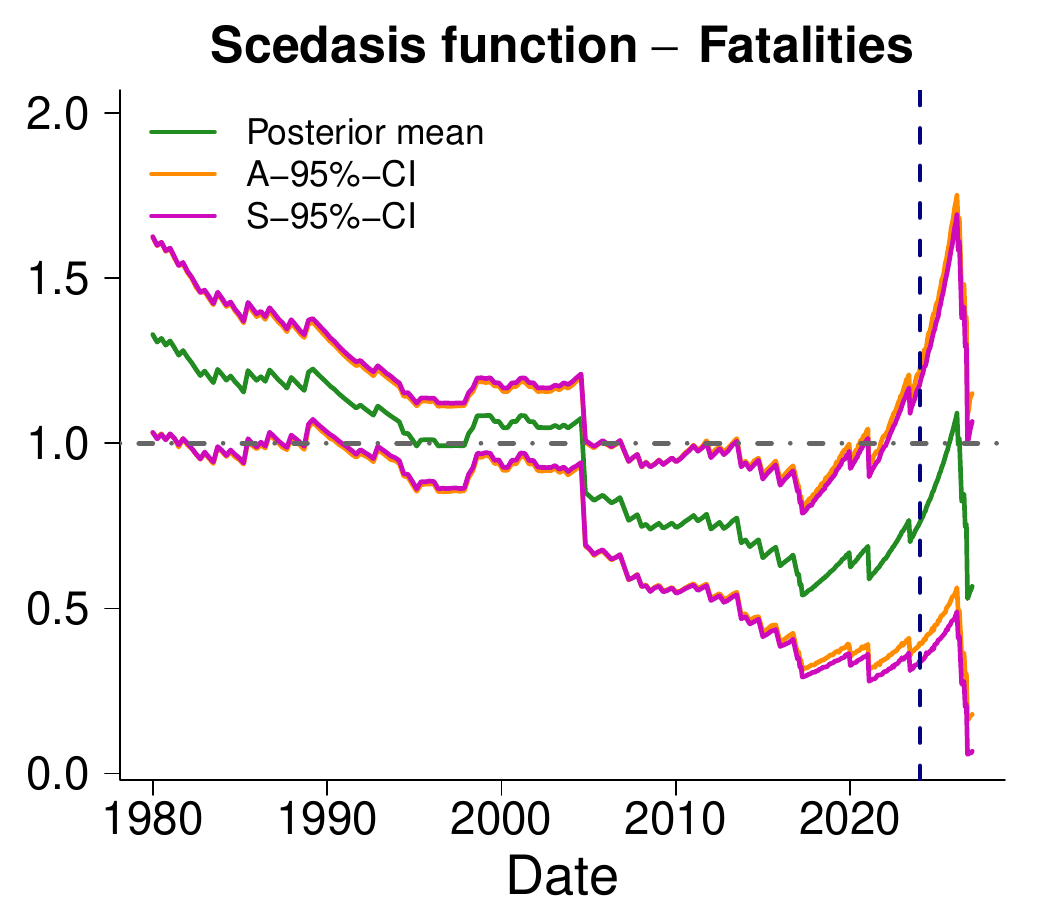}\\
\includegraphics[width=0.32\textwidth, page=27]{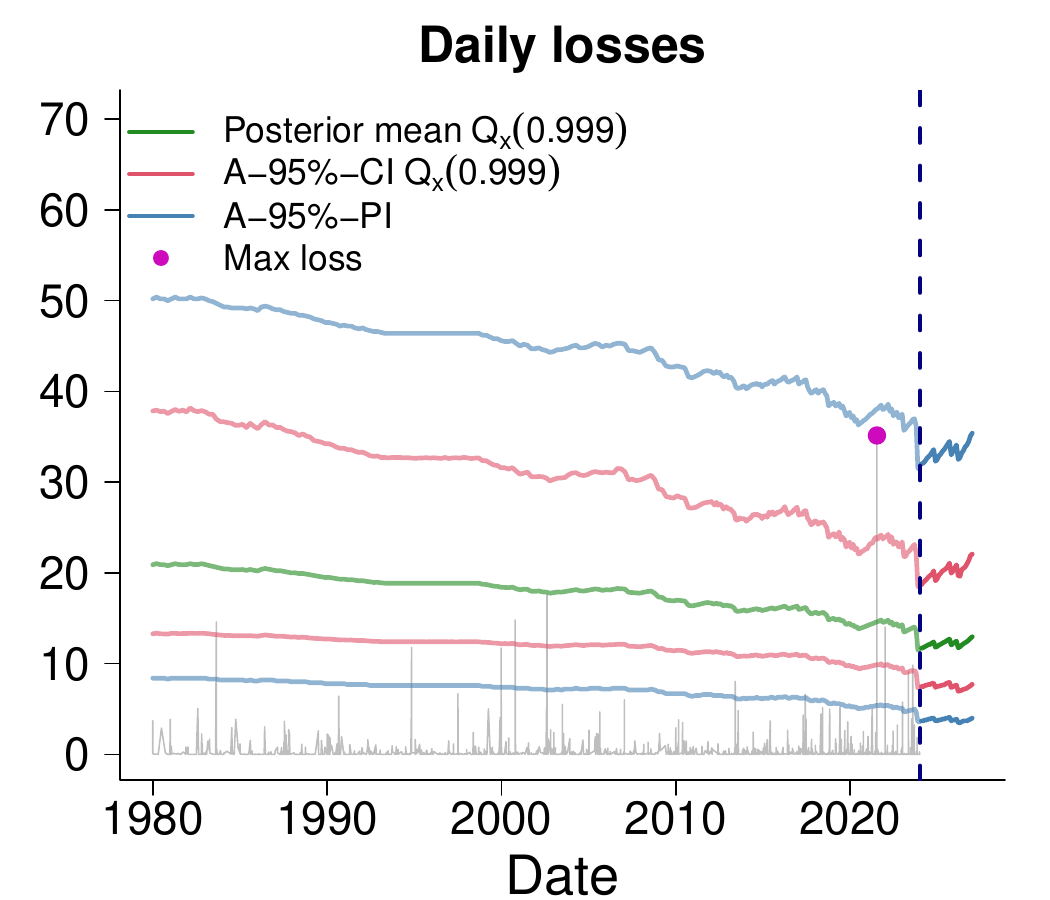}
\includegraphics[width=0.32\textwidth, page=29]{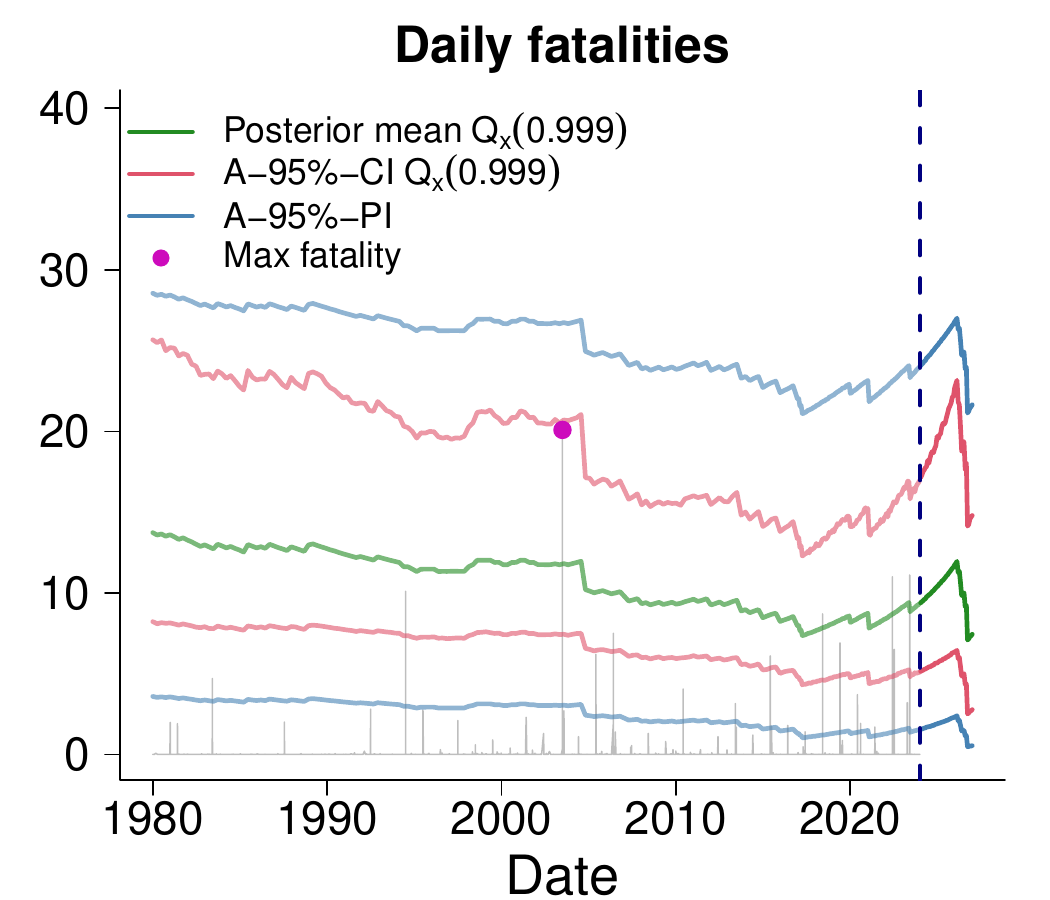}
\caption{The top panels display the scedasis posterior mean (green), along with the the A-95\%CI (orange) for losses (left panel) and fatalities (right panel). The out-of-sample period is indicated by the blue dashed line. The bottom panels show the extreme conditional quantile posterior mean (green line), accompanied by A-95\%CI (red lines) and the A-95\%PI (blue line) derived from the posterior predictive distribution for losses (left panel) and fatalities (right panel).\label{fig:sced_lossfat}}
\end{center}
\end{figure}
%

We begin the statistical analysis by testing whether the frequency of peaks changes significantly over time, using the hypothesis test described in \eqref{eq:test}. Under the null hypothesis, the scedasis function is constant, i.e., $c(x)=1$ for all $x\in[0,1]$. Table \ref{tab:test} reports the test results for both losses and fatalities. In both cases, the small $p$-values indicate strong evidence against the null hypothesis, suggesting that the frequency of extreme events varies over time. Then, we estimate the scedasis function recalling from Section \ref{sec:modnonstat} that $c(x)\approx P^*(B_n)/P(B_n)$, for all $x \in [0,1]$ and with $P(B_n)$ approximated by $\hat{p}_n= k^{-1} \sum_{i=1}^k \mathbb{I} (X_i \in B_n)$. We use a kernel-based method to estimate $P^*(B_n)$ with a fixed volume $bw=0.4$  {(losses) and 0.3 (fatalities)} for the ball $B_n$ (see \cite{dombry2023asymptotic} for further details on the estimation procedure). For all $x \in [0,1]$, we obtain then a sample of $M=20,000$ values of $c(x)$ from the posterior distribution $\Phi_n(C)=DP(C; \tau + k \mathbb{P}_n^{*} )$, where $\mathbb{P}_n^{*} =k^{-1} \sum_{i=1}^k \mathbb{I} (X_i^* \in B_n)$ and $\tau=5 \cdot bw $.
 {The top panels of} Figure \ref{fig:sced_lossfat} display  the posterior mean in green and A-95\%CI in orange of the scedasis function $\widehat{c}(x)$ for losses (left panel) and fatalities (right panel). 
The scedasis function is estimated on the observed losses (fatalities) until December 31, 2023 and evaluated on the out-of-sample period from January 1, 2024 to December 31, 2026. 
For both quantities, the value one is excluded from the credible intervals over most of the time window, confirming significant time variation in the frequency of extremes. Consistent with the exploratory patterns observed in Figure \ref{fig:bubbles}, the posterior mean of the scedasis function for losses shows a decreasing trend during the in-sample period, indicating a decline in the frequency of extreme loss events over time. In contrast, the scedasis posterior mean for fatalities remains relatively stable until around 2005, followed by a decreasing trend until approximately 2018, and then an upward trend through to 2023.
%
\begin{table}[t!]
\centering
\begin{tabular}{cccc}
\toprule
\multicolumn{4}{c}{Economic losses forecasts}\\
\toprule
Year & Posterior Mean &  A-95\%-CI &  A-95\%-PI\\
\hdashline
 2024 & \euro11.9{{0}} & [\euro 7.{5}; \euro19.{34}] &  [\euro 3.{7}0; \euro 32.{5}0] \\
2025 & \euro12.3{5} &  [\euro 7.{57};  \euro20.{57}] &  [\euro 3.{8}0; \euro 33.{7}0] \\
 2026 & \euro12.9{7} &  [\euro 7.{73}; \euro22.{08}] &  [\euro {4.00}; \euro 35.40]\\ 
 \bottomrule
\multicolumn{4}{c}{Fatalities forecasts}\\
\toprule
Year & Posterior Mean &  A-95\%-CI &  A-95\%-PI\\
\hdashline
 2024 & 10{,389} ds & [5{,649} ds; 19{,685} ds] & [1{,888} ds; 2{5,302} ds] \\
2025 & 11{,731} ds & [6{,351} ds;  22,698 ds] & [2{,319} ds; 2{6,788} ds] \\
 2026 & 7{,439} ds & [2{775} ds; 1{4,791} ds] & [~5{43} ds;  {21,646} ds]\\ 
 \bottomrule
\end{tabular}
\caption{Posterior mean of extreme quantile conditionally to time (second column), of level $\tau_E=1-1/T$, with $T$ corresponding to 31st of December of 2024, 2025 and 2026 (first column), and A-95\%CI (third column) and A-95\%PI (fourth column). Forecasts are in billion of \euro\, for losses (upper part) and deaths (ds) for fatalities (lower part).\label{tab:pfnonstat}}
\end{table}
%

We complete the analysis by discussing the results obtained from applying the proportional tail model to losses and fatalities. The bottom panels of Figure \ref{fig:sced_lossfat} show the posterior mean (green solid line) and the bounds of an A-95\%CI (red lines) for the extreme conditional quantile at level $\tau_E = 99.9\%$, computed using formula \eqref{eq:extreme_quantile_conditional}. The solid blue line denotes the bounds of the A-95\%PI obtained from the estimated predictive density of tail events conditionally to time and given that the 99.5th data percentile is exceeded, derived by formula \eqref{eq:approx_pred_covariates}. In particular,  the conditioning threshold is approximated using a Bayesian estimate. All estimates are based on observations up to December 31, 2023, with predictions made for the short-term period from January 1, 2024, to December 31, 2026. The corresponding posterior means, and A-95\%CI and A-95\%PI, are summarized in Table \ref{tab:pfnonstat}. Notably, the credible intervals are much narrower than the predictive ones, which may indicate an underestimation of risk over the prediction horizon. In fact, while the historical maximum loss falls within the predictive interval, it lies outside the credible interval.

The predictive interval also encompasses other  severe historical losses resulting from extreme climate events. Notable examples include the flood of the Var River on November 4, 1994 \citep{newsFR}, and Cyclone Martin on December 26, 1999 \citep{newsSwissRE}, both in France, which caused damages exceeding \euro11 billion. Similarly, the drought in Germany on June 1, 2005, led to 6,200 fatalities. The time-varying predictive intervals yield results consistent with the stationary case for the same extreme level: although slightly narrower, both intervals include the largest historical loss (see the third-left panel of Figure \ref{fig:uncPoTest} vs. the bottom-left panel of Figure \ref{fig:sced_lossfat}). Moreover, comparing Table \ref{tab:PFstat} and Table \ref{tab:pfnonstat}, the stationary model suggests the largest historical loss could reoccur by the end of 2024, while the nonstationary model delays this to the end of 2026—likely due to the decreasing trend in the estimated scedasis function during the in-sample period.
The year 2024 has already been marked by major hydrological disasters. 
In early June, floods and flash floods struck southern Germany and neighboring countries \cite{JunefloodDE}, and in late October, a catastrophic flash flood struck Spain’s Valencia Region. Preliminary estimates by Munich RE report US\$5 billion in damages for the former event and US\$11 billion for the latter \citep{newsMunichRE}, both occurred over short durations (one week and three days, respectively). Our A-95\%PI (converted to euros) includes both estimates, ensuring the reliability of our predictions. For instance, on  {June 6}, 2024, the A-95\%PI for extreme losses (exceeding the 99.5th percentile) ranges from approximately \euro3.8 billion to \euro33 billion, and from approximately \euro4 billion to \euro34 billion on October 30, 2024. In contrast, the corresponding credible intervals—[\euro7.64; \euro19.32] billion and [\euro7.87; \euro20.23] billion—fail to fully capture these losses, highlighting the advantage of using predictive distributions in tail risk assessment. As of now, no preliminary death toll estimates  that our model would consider as extreme are publicly available for 2024. However, the increasing trend observed in the posterior mean of fatalities in the bottom-right panel of Figure \ref{fig:sced_lossfat} aligns with 2024 being the hottest year on record across all continental regions \citep{newsFT}, and with projections indicating a net increase in heat-related deaths due to climate change, even under high adaptation scenarios \citep{masselot2025estimating}.

\section{Conclusion}\label{sec:conclusion}
We have presented a comprehensive approach for predicting future extreme economic losses and fatalities in the EU, based on historical data on climate-related hazards (e.g., floods, storms, and heatwaves) collected across the EU-27 countries over the past 43 years. Leveraging EVT, we derived a GP-based approximation of the true, unknown predictive distribution of future tail events for both losses and fatalities. By adopting a Bayesian framework, we obtained a reliable estimator of the predictive distribution, enabling robust inference and practical forecasting. The methodology starts from the simpler case of independent and homoscedastic extremes and is extended to handle non-stationary, heteroscedastic extremes over time.
The resulting toolbox is particularly valuable for conducting What-if analyses to assess hypothetical scenarios beyond observed levels, including potential worst-case outcomes. This provides a foundation for short- to medium-term predictions of tail risks, offering a critical resource for the EEA to support the design and implementation of effective adaptation strategies.

We leave for future research the extension of our framework to more general settings, including the incorporation of temporal dependence, the joint modeling of economic losses and fatalities, and the integration of covariates or outputs from weather and climate models.

\section*{Acknowledgements}
Simone A. Padoan is supported by the Bocconi Institute for Data Science and Analytics (BIDSA), Italy. 
For the data we acknowledge Stefano Ceolotto from the Euro-Mediterranean Center on Climate Change (Centro Euro-Mediterraneo sui Cambiamenti Climatici) and the European Environment Agency (EEA).

\section*{Funding}
Simone A. Padoan and Carlotta Pacifici thank for the financial support MUR–PRIN Bando 2022–prot. 20229PFAX5, financed by the European Union - Next Generation EU, Mission 4 Component 2 CUP J53D23004260001.

\section*{Data availability}
The data on losses and fatalities used in this article were provided by RiskLayer (\url{https://www.risklayer.com}) under licence, through the Catastrophe Database (CATDAT). In order to have the available data please contact RiskLayer.

\section*{Supplementary matrial}
Supplementary material is available online.

\bibliographystyle{chicago} 
\bibliography{reference}

\end{document}